\pdfoutput=1   

\documentclass{aa} 
 
\usepackage{graphicx}
\usepackage[varg]{txfonts}
\usepackage{natbib}
\bibpunct{(}{)}{;}{a}{}{,} 

\begin{document} 

 \title{A low upper mass limit for the central black hole in the late-type galaxy NGC 4414}

 \author{S.\,~Thater\inst{1} \and D.\,~Krajnovi\'{c}\inst{1} \and M.~A.\,~Bourne\inst{2} \and M.\,~Cappellari\inst{3} \and T.\,~de~Zeeuw\inst{4,5} \and E.\,~Emsellem\inst{4} \and J.\,~Magorrian \inst{3}\and R.~M.\,~McDermid\inst{6,7} \and M.\,~Sarzi\inst{8} \and G.\,~van~de~Ven\inst{9}}

 \institute{Leibniz-Institute for Astrophysics Potsdam (AIP), An der Sternwarte 16, D-14482 Potsdam, Germany, \email{sthater@aip.de} 
\and Institute of Astronomy and Kavli Institute for Cosmology, University of Cambridge, Madingley Road, Cambridge CB3 0HA, UK 
\and Sub-Department of Astrophysics, University of Oxford, Denys Wilkinson Building, Keble Road, Oxford OX1 3RH, UK 
\and ESO, Karl Schwarzschild Strasse 2, D-85748 Garching b. München, Germany \and Leiden Observatory, Leiden University, Niels Bohrweg 2, 2333 CA Leiden, Netherlands 
\and Department of Physics and Astronomy, Macquarie University, Sydney NSW 2109, Australia 
\and Australian Astronomical Observatory, PO Box 915, Sydney NSW 1670, Australia 
\and Centre for Astrophysics Research, University of Hertfordshire, College Lane, Hatfield AL10 9AB, UK 
\and Max Planck Institute for Astronomy, K\"onigstuhl 17, D-69117 Heidelberg, Germany}

 \date{Received 4 August 2016 / Accepted 4 October 2016}
 
\abstract{We present our mass estimate of the central black hole in the isolated spiral galaxy NGC 4414. Using natural guide star adaptive optics assisted observations with the Gemini Near-Infrared Integral Field Spectrometer (NIFS) and the natural seeing Gemini Multi-Object Spectrographs-North (GMOS), we derived two-dimensional stellar kinematic maps of NGC 4414 covering the central $1.5$ arcsec and $10$ arcsec, respectively, at a NIFS spatial resolution of 0.13 arcsec. The kinematic maps reveal a regular rotation pattern and a central velocity dispersion dip down to around 105 km/s. We constructed dynamical models using two different methods: Jeans anisotropic dynamical modeling and axisymmetric Schwarzschild modeling. Both modeling methods give consistent results, but we cannot constrain the lower mass limit and only measure an upper limit for the black hole mass of $M_{\mathrm{BH}}=1.56 \times 10^6\,M_{\odot}$ (at $3\,\sigma$ level) which is at least $1\sigma$ below the recent $M_{\mathrm{BH}}-\sigma_{\mathrm{e}}$ relations. Further tests with dark matter, mass-to-light ratio variation and different light models confirm that our results are not dominated by uncertainties. The derived upper limit mass is not only below the $M_{\mathrm{BH}}-\sigma_{\mathrm{e}}$ relation, but is also five times lower than the lower limit black hole mass anticipated from the resolution limit of the sphere of influence. This proves that via high quality integral field data we are now able to push black hole measurements down to at least five times less than the resolution limit.}

 
 \maketitle 
 
 \titlerunning{An upper mass limit for the central black hole in NGC 4414}
 
 \authorrunning{Thater et al.}
 
%
\section{INTRODUCTION} \label{intro}
In the last decades it has become apparent that supermassive black holes (SMBH) are embedded in the cores of most galaxies irregardless of their morphological types. By now several dozen SMBH masses ($M_{BH}$) have been determined using different measurement methods; stellar kinematics being the most commonly applied method \citep{kormendy2013,Saglia2016}. The majority of SMBH mass measurements were conducted for massive black holes in elliptical and S0 galaxies \citep[e.g., recent compilations by][]{McConnell2013,kormendy2013,Graham2016,Saglia2016,Greene2016,VandenBosch2016}. Therefore, by analyzing SMBH statistics, an intrinsic bias arises as not many low-mass late-type galaxy $M_{BH}$ have yet been measured. Recent studies of late-type galaxies include, for example, 
\cite{Greene2010,DeLorenzi2013,denBrok2015,Greene2016,Bentz2016}. Low-mass galaxies ($M_* < 10^{9.5} M_{\sun}$) were  analyzed by \cite{Reines2011,Reines2013,Seth2014,Baldassare2015}, for example.

In this study we present the black hole mass measurement of the isolated, unbarred SA(rc)c late-type galaxy NGC 4414 based on stellar kinematics. NGC~\,4414 is a flocculent spiral galaxy which shows patchy spiral arms and star formation. Except for a minor interaction with a dwarf galaxy indicated by its low surface brightness stellar shell \citep{deBlok2014}, NGC~\,4414 does not show any signs of major interactions with other galaxies \citep{Braine1997}. Therefore, its undisturbed dynamical features make NGC 4414 an excellent candidate for dynamical measurements. The inner disk of NGC~\,4414 is dominated by a stellar component having only a small dark matter contribution \citep{Vallejo2003,deBlok2014}. Based on different galaxy decomposition studies, it appears that NGC~\,4414 contains only a small and faint bulge component in its center, while having a large and massive disk.

Over 40 different distance measurements for NGC~\,4414 are available in the literature ranging from 5 to 25 Mpc. This range includes less accurate measurements from the Tully-Fisher relation.
NGC 4414 also belongs to the galaxy sample of the Hubble Space Telescope (HST) Key project \citep{Freedman2001} measuring distances based on Cepheid brightness. However, the distance measurements are still discordant and tend towards smaller distances. The measurements with the lowest errors give distances of between 16.6 and 21.1 Mpc \citep{Kanbur2003,Paturel2002}. Therefore, throughout this paper, we adopt the distance of $D=18.0\pm 3.0$ Mpc, which is the mean distance in the NASA/IPAC Extragalactic Database (NED) at the time of writing. At this distance, 1 arcsec corresponds to approximately 86.8 pc. The influence of the distance uncertainty on the black hole measurement is further discussed at the end of this paper.

Based on its observed effective stellar velocity dispersions of
$\sigma_{\mathrm{e}}\approx 110$ km/s, the $M_{BH}-\sigma_{\mathrm{e}}$ relation \citep[e.g.,][]{Ferrarese2000,Gebhardt2000,Gueltekin2009,Graham2011} predicts a central black hole mass of $8.7 \times 10^6\, M_{\sun}$ \citep[all-types;][]{Saglia2016}. While the $M_{BH}-\sigma_{\mathrm{e}}$ relation shows a very tight correlation, it is still not fully understood. A number of galaxies have been reported, which show strong deviations from the $M_{BH}-\sigma_{\mathrm{e}}$ correlation leading to the question of whether or not all different types of galaxies follow the scaling relations or show different scaling behaviors. The estimate of $M_{BH}$ within NGC~\,4414 provides a new measurement in the lower mass regime where, due to observational constraints, not many SMBHs have yet been observed. In addition, in the past, only a small number of black hole masses have been recorded for late-type Sc spiral galaxies \citep{Atkinson2005,Pastorini2007,Greene2010,Greene2016} reinforcing the need for more measurements.

In this paper, we present optical and adaptive optics-assisted near-infrared integral-field spectroscopic data for NGC 4414, in order to study the stellar kinematics in the vicinity of its central black hole. In Section~\ref{data}, we describe our observational data and in Section~\ref{kinematics} the extraction of the stellar kinematics from the GMOS and NIFS integral-field spectroscopic data. In addition to the kinematics, we combine high-resolution HST and Sloan Digital Sky Survey (SDSS) data to model the stellar surface brightness and thus examine the stellar brightness density of NGC~\,4414. In Section~\ref{dynamical modelling} we present the dynamical models which we constructed using two different methods: 1) Jeans Axisymmetric Modeling \citep{Cappellari2008} and 2) Schwarzschild’s orbit superposition method \citep{Schwarzschild1979}. We analyze our assumptions for the dynamical modeling and discuss our results in the context of the $M_{\mathrm{BH}}$-host galaxy relationships in Section~\ref{discussion}, and finally conclude in Section~\ref{conclusions}.
\begin{table}
\caption{Basic properties of NGC~\,4414 taken from the literature.}
\centering
\begin{tabular}{lcc}
\hline\hline
Property &   &  Reference \\
\hline
Morphological type     &  SA(rc)c   & 1   \\
Distance [Mpc]   &  $18 \pm 3.0$  & 2    \\
Inclination [ $^{\circ}$] & 55 & 3 \\
Bulge effective radius [arcsec]    & $3.9 \pm 1.4$ & 4    \\
$\sigma_e$ [km/s]    &  $115.5 \pm 3$  & 5  \\

\hline
\end{tabular}
\tablefoot{1 - \citet{deVaucouleurs1991rc3}, 2 - Mean distance from NED, 3 - \citet{Vallejo2002}, 4 - \citet{Fisher2009}, 5 - This work.}
\label{properties}
\end{table}

\section{OBJECT SELECTION, OBSERVATIONS \& DATA REDUCTION} \label{data}
We used the NIFS and GMOS ground-based integral-field units (IFU) to obtain stellar kinematics in two spatial dimensions allowing better constraints on the black hole mass estimate. High-resolution NIFS data is essential for a precise measurement of the stellar motions in the vicinity of the central black hole, while the large-scale GMOS data constrains the global stellar mass-to-light ratio (M/L) and the stellar orbital distribution. 

\subsection{Object Selection}
In order to achieve the best possible resolution to probe the vicinity of the central black hole, we wanted to utilize adaptive optics in combination with a natural guide star (NGS). Therefore, we undertook a careful study to identify possible targets with bright nearby reference stars by cross-correlating the all-sky 2MASS  point and extended source catalogues \citep{Skrutskie2006}. We searched for feasible NGSs close to all galaxies which fulfilled certain criteria: 1) being a northern object since we conducted our observations at the GEMINI observatory; 2) having an available stellar velocity dispersion measurement allowing prediction of the black hole mass based on the $M_{BH}-\sigma_{\mathrm{e}}$ relation \citep{Ferrarese2005}, that was relevant at that time; 3) from the stellar velocity dispersion and the predicted black hole mass we calculated the sphere of influence of the central black hole given by $r_{\mathrm{infl}}=G\,M_{BH}/\sigma^2$ where G is the gravitational constant and $\sigma_e$ the stellar velocity dispersion of the host bulge and only considered objects with $r_{\mathrm{infl}}>0.06''$ reaching the diffraction limit of Gemini at 2.3 micron \citep{Krajnovic2009a,Cappellari2010}; 4) the existence of high-resolution HST imaging data. These NGS specifications yielded a sample of six galaxies which were observable at the time of the data acquisition from which NGC 4414 was the only unbarred spiral. 

\subsection{GEMINI NIFS data} \label{NIFS data}
\begin{figure}
\centering
   \includegraphics[width=0.8\hsize]{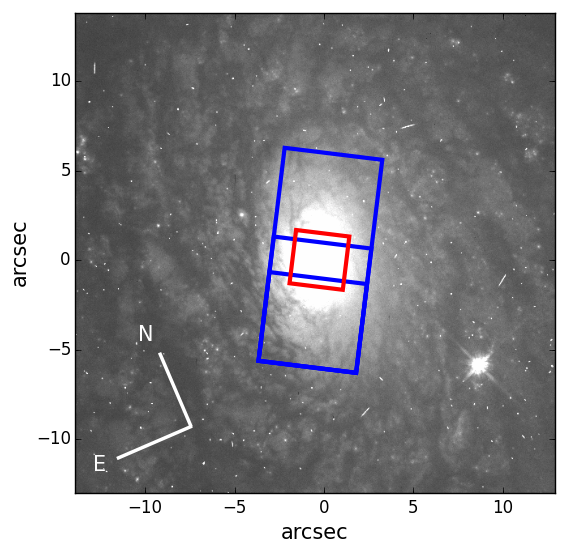}
      \caption{Schematic overplot of the NIFS (red) and GMOS (blue) field-of-view on the WFPC2 F606 image. The kinematic data only map the very core of NGC~\,4414 which is not polluted by dust. The orientation of the NIFS data is $150^\circ$ and that of the GMOS data is $60^\circ$ counterclockwise from north. The star in the south-western part of the bulge served as a natural guide star for our observations.}
         \label{nifsgmos}
\end{figure}

Using the NIFS instrument \citep[][]{McGregor2003}, NGC~\,4414 was observed in May and June 2007  at the 8.1m Gemini North telescope under the science program GN-2007A-Q-45. In order to improve the natural seeing correction, NIFS operated by using adaptive optics with natural guide star assisted mode. Integral-field spectroscopy in combination with adaptive optics reduces wavefront distortions of the Earth’s atmosphere and thus increases the resolving power of the telescope.  Figure~\ref{nifsgmos} shows the star, which served as the natural guide star for our observations. Projected on the sky it is around $10''$ away from the center of our target. The observation was conducted in the K-band to reduce dust contamination and covers a field of view (FOV) of approximately $3'' \times 3''$ centered on the core of NGC~\,4414 and covering its bulge (Fig.~\ref{nifsgmos}). In total, NGC~\,4414 was observed for $34 \times 600$ seconds ($\sim 5.5$ hours) with NIFS.

The NIFS observations were reduced using the Gemini NIFS reduction routines which are provided in IRAF\footnote{http://www.gemini.edu/sciops/instruments/nifs/data-format-and-reduction}. The data reduction includes bias and sky subtraction, flatfield calibration, interpolation over bad pixels, cosmic-ray removal, spatial rectification and wavelength and flux calibration with arc lamp exposures. 
After the data reduction, we merged eleven individual science frames into a final data cube to increase the flux of the spaxels and thus the signal-to-noise ratio (S/N). The science frames had to be re-centered to a chosen reference frame by determining their relative shifts to each other and calibrated to the same wavelength range and pixel sampling. According to the method presented in \citet{Krajnovic2009a}, the re-centering was performed by comparing and re-aligning the science frame's isophotes. The isophotes could not always be matched perfectly as in some cases the outer and inner isophotes were not concentric. This inconsistency can result from uncertainties in the adaptive optics correction. While the inner isophotes are more strongly influenced by the point spread function (PSF), the outer isophotes contain less flux and are more affected by statistical uncertainties. Therefore, we used a compromise between the geometrically more robust outer isophotes and the more flux-significant inner isophotes to deduce the relative shifts. 
Using the determined shifts, all of the frames were aligned to the reference frame. Finally, the frames were merged with a sigma-clipping pixel reject algorithm to create the final data cube. A new square pixel grid of $0.05''$ scale was defined, individual science frames were interpolated to this grid and the flux values of the final data cube calculated as the median flux values of the single data frames as in \citet{Krajnovic2009a}.

Before extracting kinematics from a data cube a high S/N has to be guaranteed; typically S/N $\geq 40$ \citep{vanderMarel1993,Bender1994}, as the higher-order moments of the line-of-sight velocity distribution (LOSVD) are very sensitive to noise effects. To keep a roughly constant, high S/N, we used the Voronoi adaptive binning technique\footnote{http://purl.org/cappellari/software} introduced by \citet{Cappellari2003}. As the size of the Voronoi bins varies according to the local S/N, a high spatial resolution can be retained in the central pixels. An initial estimate of the noise N of the unbinned spectra was determined by smoothing each galaxy spectrum over 30 pixels and then taking the standard deviation
of the residual between the respective galaxy spectrum and the smoothed spectrum. According to our initial S/N estimate, the critical S/N threshold was chosen to be 60. Using the Voronoi binning method, adjacent pixels of our data were binned together into 1546 bins. The single spectra of the binned pixels were co-added to provide a S/N $\geq$ 60.
Then, we computed the residual-noise (rN) of each bin as the standard deviation of the difference between the observed galaxy spectrum and the kinematic model (see Sect.~\ref{kinematics}). The final signal-to-residual-noise S/rN lies between 35 and 85, with lower values at the edges of the FOV.

\subsection{GEMINI GMOS-N data} \label{GMOS data}
Parallel IFU observations were conducted with the GMOS instrument \citep[][]{Allington-Smith2002} at the Gemini Observatory to cover the complete bulge of NGC~\,4414 (science program GN-2007A-Q-45). The GMOS observations were performed with the $B600$ grating in the $g\_G0301$ filter and with two pointings keeping the target nucleus in both. The combined FOV covers $5.5'' \times 12''$ which enables us to further constrain the bulge of our target. In total, NGC~\,4414 was observed $5 \times 1800$ seconds (2.5 hours) at two different wavelengths (475 and 483 nm).  Figure~\ref{nifsgmos} displays the FOV and the orientation of the NIFS and GMOS data relative to each other and to the extent of the galaxy.
Different routines of the IRAF package (see footnote 1) were used for the data reduction of the GMOS data, which consisted of bias subtraction, correction of the spectrum-pixel-association, flat-fielding, wavelength-calibration, and cosmic ray removal. The wavelength-calibration was carried out by comparing the GMOS spectral lines with reference spectral lines of a copper-argon lamp. In addition, a bad pixel correction was applied. One of the blue fiber bundles of the GMOS spectrograph had reduced flux passage resulting in two columns of bad pixels in the science frames. Depending on whether the bad columns lay inside or outside of the observed galaxy nucleus, two different corrections were applied. Due to symmetry, bad pixels inside the galaxy nucleus were corrected by mirroring the flux level of the opposite side of the galaxy nucleus. On the other hand, outside the galaxy nucleus, bad pixels were corrected by interpolating the horizontally adjacent pixels. This correction only affects the absolute flux level of the bad pixels, but does not influence the absorption lines, and, therefore, does not alter the kinematics. In total, nine science frames were merged with the method described in Sect.~\ref{NIFS data} to create a final GMOS data cube. As the central region of NGC~\,4414 was mainly probed with the better resolved NIFS data, we Voronoi binned adjacent spaxels together to achieve a S/N $\ge 60$ for each spaxel. The final S/rN of the GMOS data lies between 40 and 100.

\begin{figure*}[!htb]
\centering
   \includegraphics[width=\hsize]{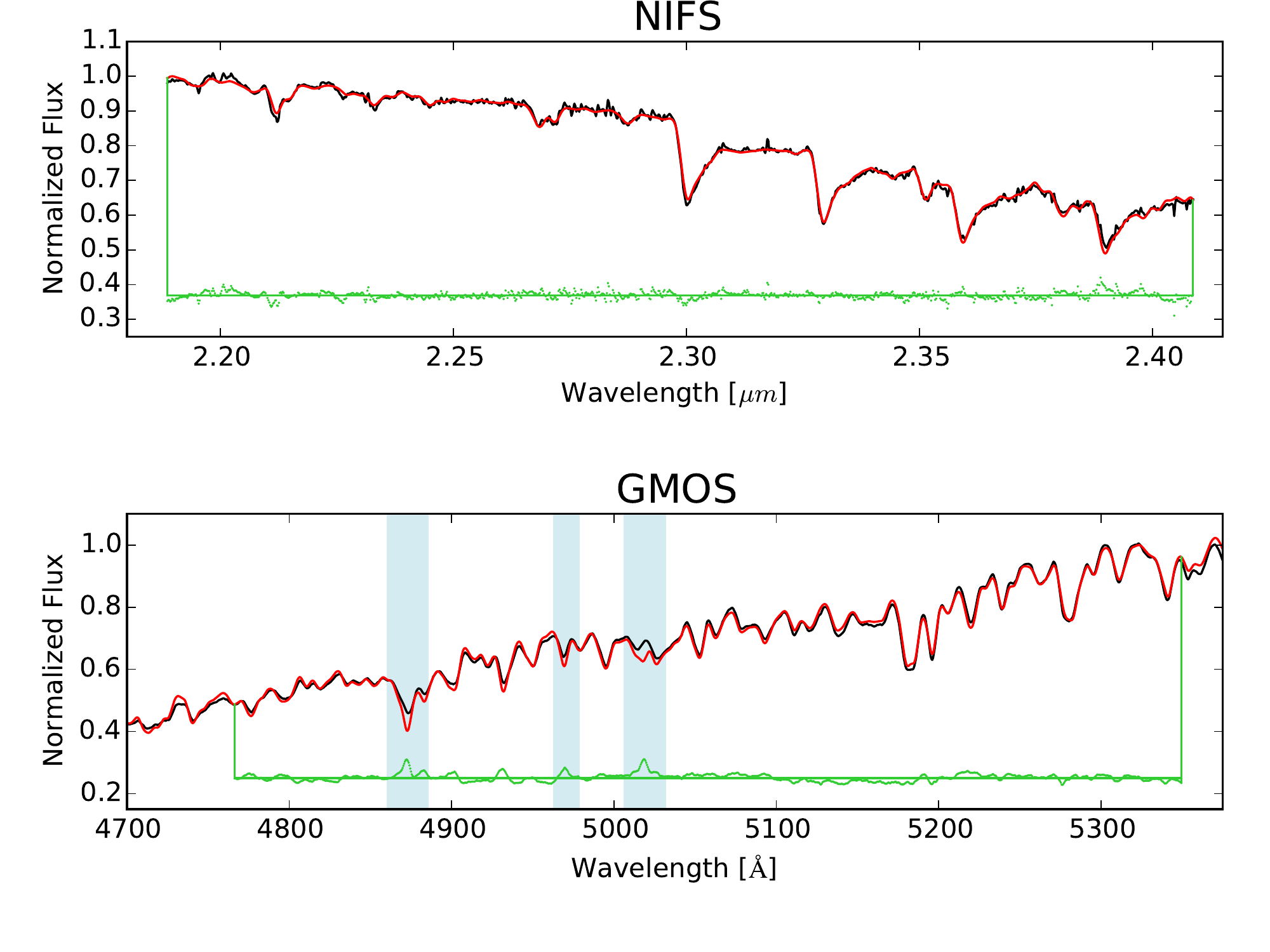}
      \caption{Optimal template for NIFS (top) and GMOS (bottom) observations. The black line represents the total galaxy spectrum, which is composed of the spectra of all single spatial pixels, the red line shows the optimal template, which is fitted to the total galaxy spectrum. The fitting residuals are presented as green dots spread around the residual=0-line (green solid line), where both are shifted upwards by an arbitrary amount. The shaded regions contain emission lines and are not included in the fit.}
         \label{optimaltemplate}
\end{figure*}

\subsection{Imaging data} \label{Imaging}
In order to construct dynamical models, appropriate imaging data of NGC~\,4414 is required to measure the surface brightness and determine the gravitational potential of the stellar component. The imaging data must have a large FOV to cover the light of the whole galaxy and a sufficient spatial resolution in the very central regions where the central mass dominates the galaxy potential. Therefore, we retrieved Wide Field and Planetary Camera 2 (WFPC2) imaging data in F606W from the ESA Hubble Science Archive, which generates automatically reduced and calibrated data. Except for F606W, all WFPC2 images of NGC~\,4414 were saturated in the center and were therefore not usable for our modeling. The prior data calibration process involved the steps of masking bad pixels, performing an analog-to-digital (A/D) correction and the correction of bias, dark and shutter shading (WFPC2 Handbook\footnote{http://documents.stsci.edu/hst/wfpc2/documents/\\handbooks/dhb/wfpc2\_dhb.pdf}). Because of the central saturation of most of the WFPC2 images, it was not possible to remove cosmic rays by comparing the different images. That is why we corrected the images for cosmic rays by determining pixels (at least five pixels beyond the center) that had a large count gradient towards their neighbors and masked these pixels prior to the fitting of the light distribution (Sect.~\ref{mass model}). We also masked pixels which were significantly affected by dust extinction. These pixels were determined by the method described in Sect.~\ref{SDSSdust}.

In order to fulfill the large FOV criteria, we used ground-based SDSS imaging data in the g-, r- and i-band as our second source. We used the Montage-based online tool Image Mosaic Service\footnote{http://montage.ipac.caltech.edu/} provided by NASA and Caltech to create a $0.2 \times 0.2$ square degrees mosaic centered on NGC~\,4414. Image Mosaic Service uses the individual SDSS images and overlaps them by preserving the fluxes and astrometry.

Dust can have a significant effect on the accuracy and reliability of photometric models as it alters the apparent shape of the galaxy and dims the light due to the wavelength dependent extinction. The images of NGC~\,4414 indicate large dust and gas patterns in the disk of the galaxy, which obscure the attained light of the galaxy in the SDSS r-band images such that the actual surface brightness is not observable. Using a method described in \citet{Cappellari2002b} we corrected the SDSS r-band large FOV image for these extinction effects in order to be able to fit the underlying surface brightness profile. The details of the dust correction are outlined in \ref{SDSSdust}. The largest correction was around 25\%\ of the measured flux (see Fig.~\ref{dustcorrection2}). After the correction, the disk region of NGC~\,4414 shows a more homogeneous flux distribution than before with a larger dust correction on the eastern side of the galaxy.

\section{STELLAR KINEMATICS} \label{kinematics}
\subsection{Method} \label{kinematics method}
\begin{figure*}
\centering
   \includegraphics[width=\hsize]{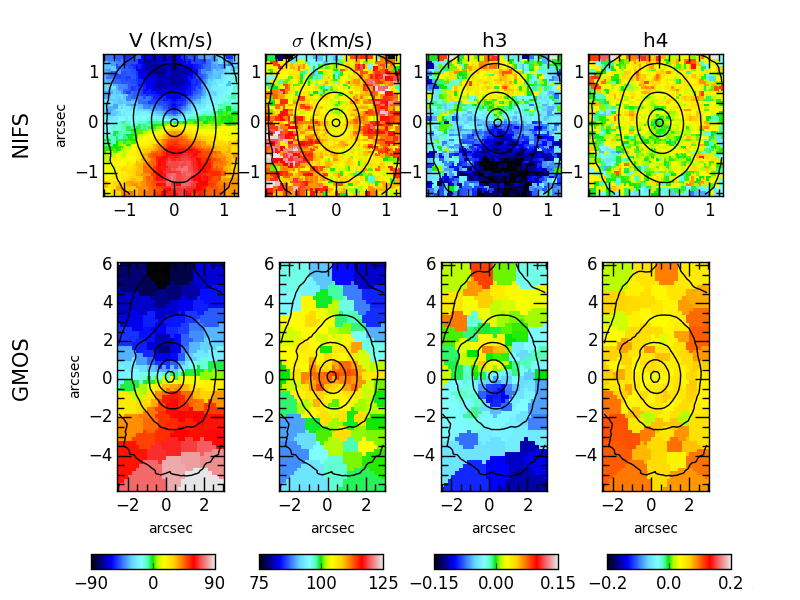}
      \caption{Kinematic maps of NGC~\,4414 derived from NIFS (top) and GMOS (bottom) observations. From left to right the maps show: the mean velocity V, the velocity dispersion $\sigma$ and the Gauss-Hermite moments: h$_3$ and h$_4$. The maps are aligned such that the northern side of the galaxy is at the top. }
         \label{kinematic maps}
\end{figure*}
The observed spectrum of a galaxy is the convolution of the integrated spectrum of its stellar population with the instrumental broadening and the LOSVD. In order to extract the kinematics from the galaxy absorption line spectra, we used the penalized Pixel-fitting method  \citep[pPXF (see footnote 2);][]{Cappellari2004}. pPXF fits the observed spectrum by convolving a template spectrum with the LOSVD which is parametrized by the mean velocity V, the velocity dispersion $\sigma$ and Gauss-Hermite polynomials \citep{Gerhard1993,vanderMarel1993}. The success of pPXF depends critically on the provision of a good set of template stellar spectra that match the galaxy spectrum as closely as possible. In order to avoid a template mismatch, an `optimal template' is created as a linear combination of spectra taken from a stellar template library covering the observed wavelength region. The stellar templates used for GMOS and NIFS were the Medium resolution INT Library of Empirical Spectra \citep[MILES; ][]{Sanchez-Blazquez2006} and the \citet{Winge2009} compendium of Gemini Near Infrared Spectrograph (GNIRS) and NIFS observed stars, respectively.

We applied the same general procedure to both data sets: assuming that the whole galaxy consists of the same stellar population, we first determined the optimal template by fitting the stellar template library to a composite spectrum which was created by adding all spectra of the galaxy data cube. During the later application of pPXF, this optimal template was fitted to each bin of the data cube in order to determine the LOSVD of each spatial bin. We specified the LOSVD with the V, $\sigma, h_3$ and $h_4$ moments and included a fourth degree additive polynomial in order to correct the shape of the underlying continuum. From the pPXF fitting we obtained the values of the moments for each spatial bin. We then checked each spectrum visually for template mismatch and for the quality of the fitting, but no peculiarities were found.

The uncertainties of the stellar kinematics were determined by using Monte Carlo simulations. Therefore, we applied the following procedure to the spectrum of each Voronoi bin: we determined the standard deviation between the spectrum and the pPXF fit and perturbed the spectrum 100 times by adding an appropriate random Gaussian noise at the order of the standard deviation. For each of the 100 realizations, the LOSVD was determined. Finally, we took the standard deviation of the derived LOSVD moments.

The following two sections explain the particular specifications for the two data sets and the results from the kinematic fits are presented in Sect.~\ref{kinematic results}.

\subsubsection{NIFS Specifics} \label{NIFS kinematics}

As NIFS provides spectra in the near-infrared regime, a stellar template catalog in the K-band is required. Therefore, we used two stellar template libraries\footnote{http://www.gemini.edu/sciops/instruments/nearir-resources/spectral-templates} by \citet{Winge2009} which consist of G-, K- and M-stars with spectra centered at $2.2\, \mu$m and a spectral resolution of approximately $3.2\,$\AA. One template archive was observed with GNIRS (23 stars), the other with NIFS (31 stars). In order to work with the GNIRS stellar library it was necessary to take the different instrumental resolutions into account. By fitting ten characteristic lines of the sky observation, we determined the spectral resolution of NIFS to be $\sigma_{\mathrm{NIFS}}= 3.2\,$\AA\, compared to GNIRS with $\sigma_{\mathrm{GNIRS}}= 2.9\,$\AA. Therefore, we had to convolve the GNIRS templates with the quadratic difference in resolution. Furthermore, we used the spectrum to the interval shown in Fig.~\ref{optimaltemplate} to mitigate edge effects and possible contamination from emission lines. The most significant features in the NIFS spectra are the four CO absorption lines which are well fitted by pPXF.  

\begin{figure}
\centering
   \includegraphics[width=\hsize]{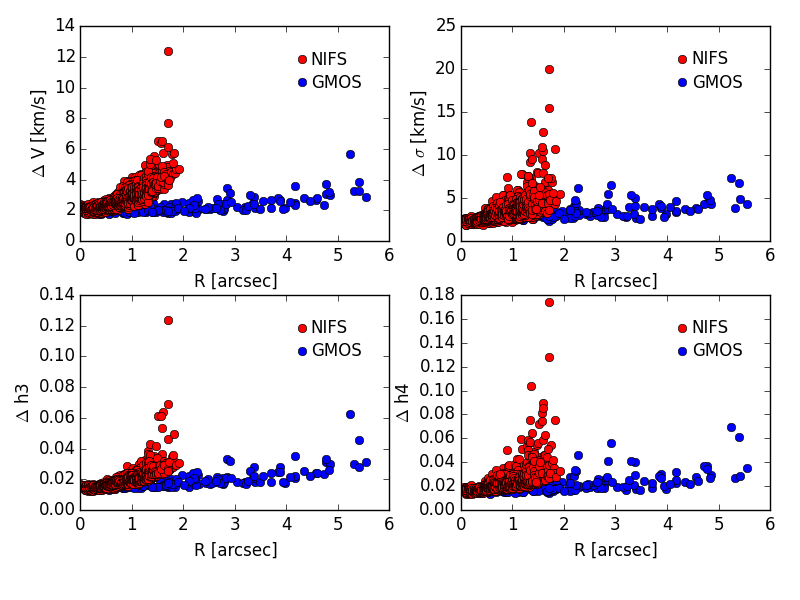}
      \caption{Radial distribution of kinematic errors of the NIFS (red) and GMOS (blue) observation. Shown are the errors of the mean velocity $\Delta\, V$, velocity dispersion $\Delta \, \sigma$, $\Delta\, h_3$ and $\Delta \,h4$. The errors are uniformly low in the central regions and increase towards the edges of the maps.
              }
         \label{kinematic errors}
\end{figure}

\subsubsection{GMOS Specifics} \label{GMOS kinematics}
For the optical GMOS data, we used the stellar templates from the MILES library, which covers a wavelength range from 3525 \AA\, to 7500 \AA\, and consists of 985 stars in total. Similarly to the NIFS preparations, the difference between the stellar template resolution and the instrumental resolution has to be taken into account when comparing the different data sets. The instrumental resolution of GMOS was derived by applying the pPXF routine on an extracted twilight exposure and measuring the mean velocity dispersion by using a solar template. Thus, we achieved the spectral resolutions of $\sigma_{\mathrm{GMOS}}=2.16$ \AA\, compared to  $\sigma_{\mathrm{MILES}}=2.5$ \AA\, \citep{Falcon-Barroso2011}. As both instrumental resolutions are similar, we did not convolve the GMOS spectra to lower resolution in order to retain important kinematic information. We analyzed the effect of not taking the convolution into account by using a well-defined sub-sample (in total 52 spectra) of the Indo-US stellar library \citep{Valdes2004} which covers a wavelength range of 3460 to 9464 \AA\, and a spectral resolution of $\sigma_{\mathrm{Indo-US}}=1.35$~\AA\, \citep{Beifiori2011}. We extracted the kinematics of GMOS based on the Indo-US library, but this time taking the difference in resolution into account. The resulting kinematic maps show the same general features and trends as the kinematic maps which are based on the MILES template and without accounting for convolution. However, the comparison revealed systematic offsets in the even LOSVD moments of  $\sigma \approx +8$ km/s and $h_4 \approx 0.02$. These offsets are in the order of a factor of 3 and 2 times the corresponding errors for the central and 2 and 0.5 times the errors for the outer spaxels.
Differences between the kinematic results could be inferred from a template mismatch in the Indo-US library since we build our optimal template from only 52 stellar spectra. Therefore, we decided to use the MILES spectral library in the further analysis, whilst keeping the offsets in the even velocity moments as additional uncertainty.

\subsection{Kinematic Results} \label{kinematic results}
The two-dimensional kinematic maps for NIFS (top) and GMOS (bottom) are presented in Fig.~\ref{kinematic maps}. The two panels on the left show the rotational velocity of NGC 4414 for the NIFS and GMOS observations. Both observations are consistent with each other and show regular rotation patterns without any major asymmetries. The northern part of NGC~\,4414 moves towards us, while the southern part is the receding side. After subtraction of the systemic velocity, the extreme values are approximately $\pm 77 $ km/s for the NIFS observation and $\pm 90$ km/s for GMOS. The velocity dispersion map shows an elongated dip down to approximately 105 km/s in the central region of the NIFS data which is orientated along the major axis. This dip is not distinctly visible in the GMOS data due to the limited spatial resolution and the large Voronoi bins in the center. Instead, the GMOS data shows a dumbbell-shaped central increase with peculiar lobes along the minor axis of approximately 114 km/s. Both the extended maximum of the GMOS velocity dispersion and the elongated minimum of the NIFS velocity dispersion coincide with the photometric center of NGC~\,4414. 
The third Gauss-Hermite moment h$_3$, which is loosely related to the skewness of the distribution, is strongly anti-correlated to the rotational velocity. The h$_4$ map, which relates to the kurtosis, shows a dip in the central $2\arcsec$ of the GMOS data, while it is uniformly distributed over the NIFS h$_4$ map (with an underlying gradient from blue to red towards the northern side).

The radial distribution of the errors derived from the Monte Carlo simulations are illustrated in Fig.~\ref{kinematic errors}. The NIFS errors are uniformly low in the central $1\arcsec$, but increase sharply towards the edges of the map and thus follow the S/rN of the observation. A similar behavior is seen in the GMOS errors. However, the GMOS errors are lower as the spectra were binned to a higher S/rN. Apart from the central region of
the velocity dispersion map, the overall comparison between the NIFS and GMOS data shows good agreement. This allows a combination of both data sets in the latter sections of this paper.

\subsection{Spatial resolution of NIFS and GMOS} \label{resolution}
A key parameter for characterizing the quality of the data is the effective spatial resolution expressed as the width of the PSF of the reconstructed unbinned data cube. A precise measurement is not only important to estimate the seeing conditions during the observations but also to determine how far the dynamics in the center of the galaxy can be probed. In order to measure the NIFS PSF, we convolved the HST data with the sum of two concentric circular two-dimensional Gaussians such that it matched the NIFS data \citep[as done e.g., by][]{McDermid2006,Davies2008,Krajnovic2009a}. The GMOS PSF was determined by convolving the HST data with only a single Gaussian such that it matched the GMOS data. Both PSFs are parametrized by the full width at half maximum ${\rm FWHM}_{PSF}$ of each Gaussian component and their relative fluxes. Details of the fitting and the (double) Gaussian fits are given in Sect.~\ref{NIFS GMOS PSF} and the parameters are given in Table~\ref{PSF parameters}. The NIFS PSF is comparable to measurements from other papers which use laser guide star adaptive optics \citep{Krajnovic2009a,Walsh2015}. In addition, we used the narrow component of the NIFS PSF to determine the Strehl ratio which yields $30.8$\% (details are given in Appendix~\ref{strehlratio}).

\begin{table}
\caption{Parameters of the double Gaussian fits for the NIFS and GMOS PSF. Given is the full width at half maximum ${\rm fwhm}_{PSF}$ of the two Gaussians and the relative flux $f_1$ of the first Gaussian.}
\centering
\begin{tabular}{ccc}
\hline\hline
Data & NIFS  &  GMOS \\
 & (arcsec)  &  (arcsec) \\
\hline
${\rm fwhm}_1$ & 0.126 & 1.05\\
${\rm fwhm}_2$ & 1.19 & -\\
$f_1$ & 0.7 & 1\\
\hline
\end{tabular}
\label{PSF parameters}
\end{table}
 \begin{figure}
\centering
\includegraphics[width=0.9\hsize]{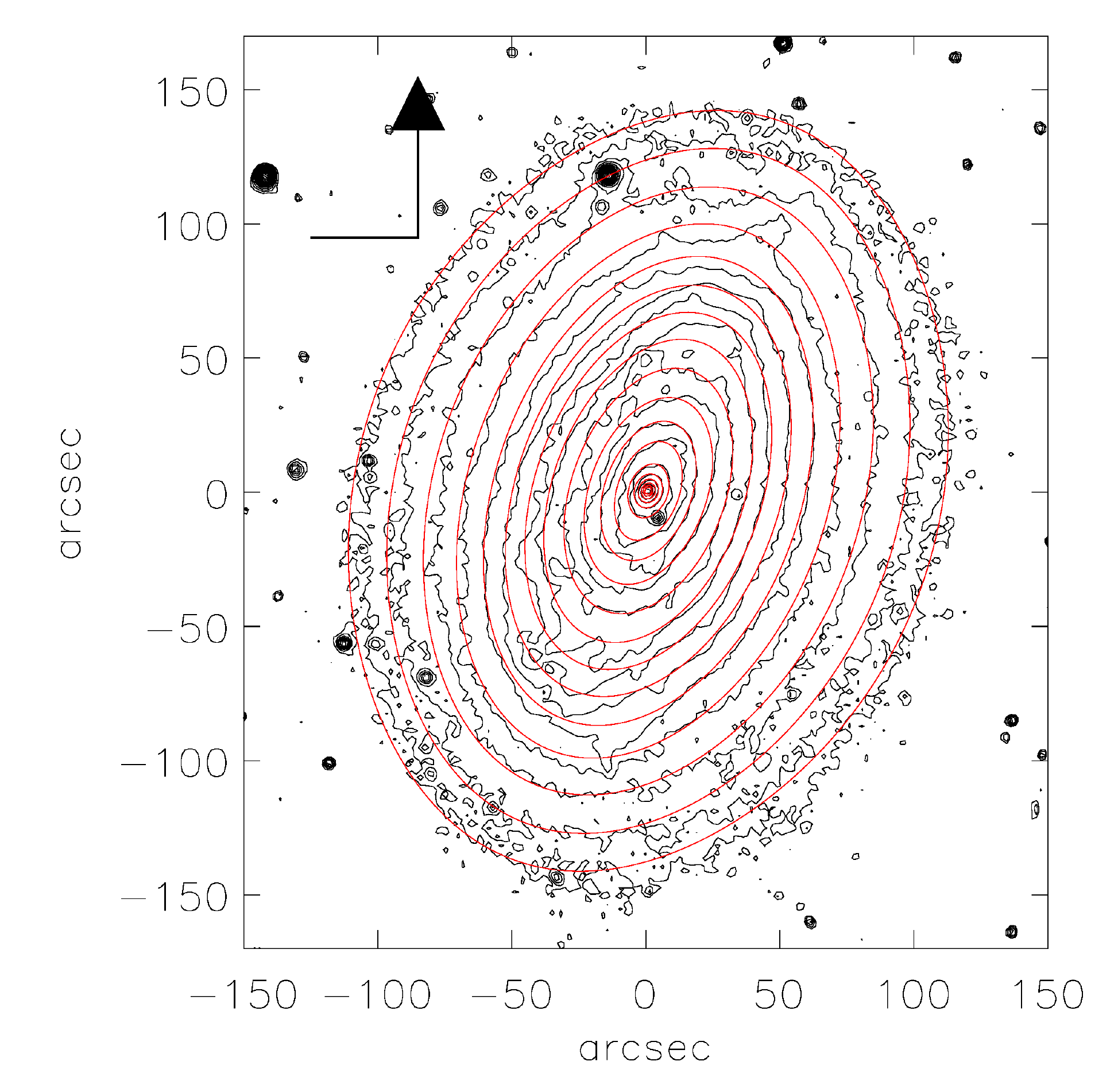}
   \includegraphics[width=1.\hsize]{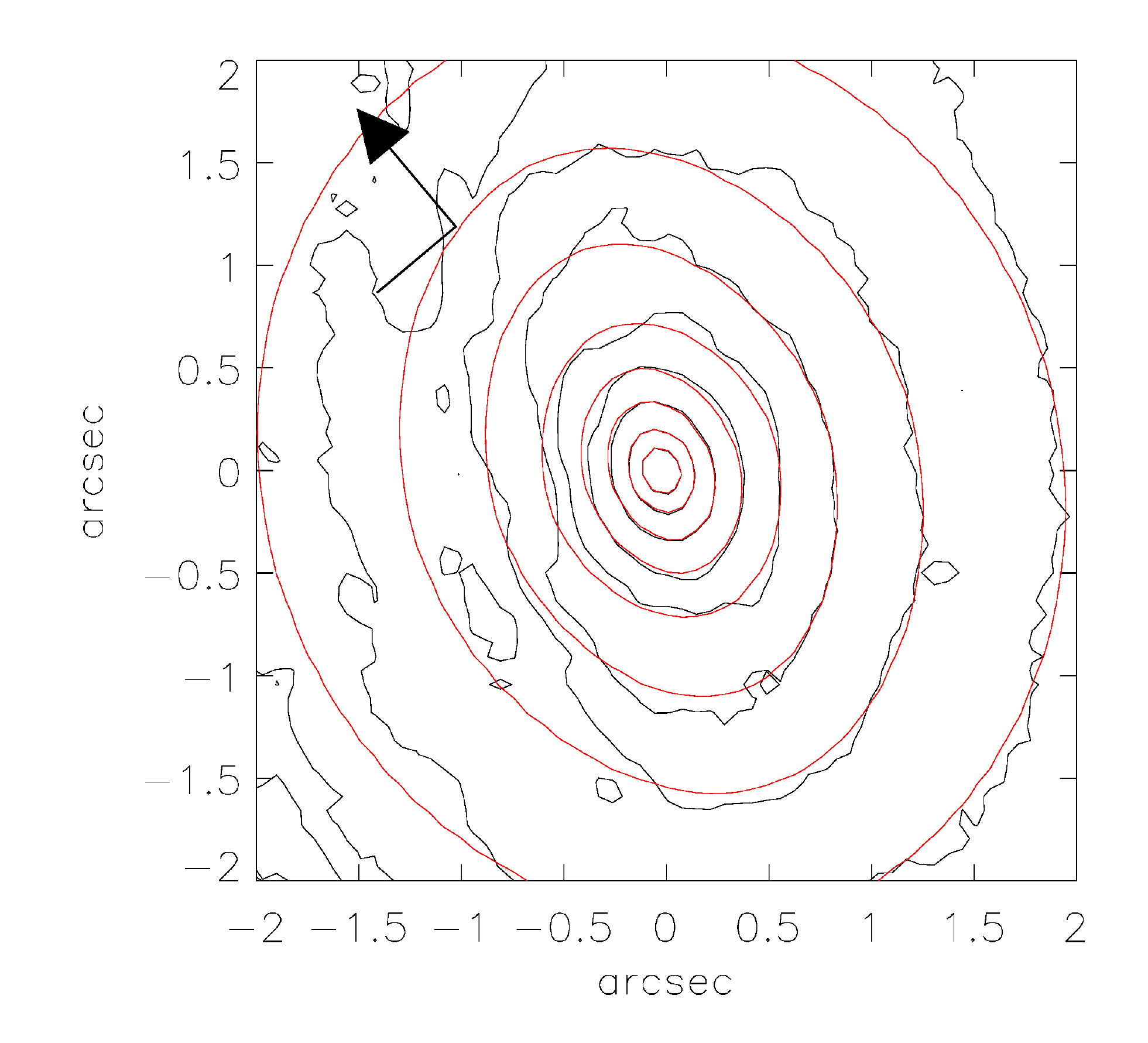}
      \caption{Comparison between the r-band isophotes of NGC~\,4414 (black) and the MGE surface brightness model (red) from the combination of the HST F606W and the SDSS r-band image. The upper panel shows a wide-field view with isophotes from the SDSS r-band image, the lower panel shows a magnification to the central $4 \times 4''$ region based on the HST F606W isophotes.  Foreground stars and strong dust patterns were masked during the MGE fit. The upper panel is orientated such that north is at the top and east is on the left of the image, the lower panel is orientated as in Fig.~\ref{nifsgmos}.}
         \label{mgefit}
\end{figure}
\subsection{The effective stellar velocity dispersion $\sigma_{\mathrm{e}}$} \label{bulge dispersion}
A prediction for $M_{BH}$ in NGC~\,4414 can be achieved by inserting its effective stellar velocity dispersion $\sigma_{\mathrm{e}}$ into the $M_{BH}-\sigma_{\mathrm{e}}$ relation. The effective stellar velocity dispersion is the integrated velocity dispersion within one bulge effective radius and can be determined from our spectral data. Unfortunately, the GMOS data does not completely cover the bulge effective radius $R_{\mathrm{e}}$ of NGC~\,4414, which is approximately $3.9 \pm 1.4''$ \citep{Fisher2009}. Consequently, it was necessary to extrapolate the measured integrated velocity dispersion towards $\sigma_{\mathrm{e}}$. We considered two different methods for determining $\sigma_{\mathrm{e}}$: 1) extrapolating $\sigma$ to the desired value on the basis of typical velocity dispersion profiles and 2) extracting a velocity dispersion map from Jeans Anisotropic Modelling \citep[JAM;][(see footnote 2)]{Cappellari2008} models.

Based on galaxy velocity dispersion profiles of the CALIFA sample \citep{Sanchez2012}, \cite{Falcon-Barroso2016b} provide an extrapolation of $\sigma_{\mathrm{e}}$ for late-type galaxies and extend the work of \citet{Cappellari2006} on early-type galaxies. They show that the velocity dispersion profile for galaxies follows a power-law which has the form $(\sigma_{\mathrm{R}}/\sigma_{\mathrm{e}})=(R/R_{\mathrm{e}})^{\alpha}$ where $\sigma_R$ denotes the velocity dispersion at a given radius R. The exponent $\alpha$ is positive for late-type galaxies ($\alpha_{\mathrm{LTG}}=0.077$) and negative for early-type galaxies ($\alpha_{\mathrm{ETG}}=-0.055$). While NGC 4414 is a late-type galaxy, its GMOS $\sigma$ map shows a decreasing trend with radius which is why we also tested the power-law with a negative exponent for our data.
In order to measure $\sigma_R$ from the GMOS data, we first co-added the spectra of each spaxel within an elliptical aperture of radius $2 \arcsec$. The resulting spectrum equals a spectrum that would have been observed with a single aperture having a semi-major axis of $R_1=2''\approx 1/2 \, R_{\mathrm{e}}$. Applying pPXF to the composed spectrum provides $\sigma_{R,1}=$121.5 km/s. We used that value in the power-law equations and obtained $\sigma_{\mathrm{e,LTG}}=$ 127.9 $\pm$4 km/s and $\sigma_{\mathrm{e,ETG}}=$ 117.1 $\pm$ 3 km/s. Using $\alpha=-0.066$ \citep{Cappellari2006} results in $\sigma_{\mathrm{e,ETG}}=$ 116.3 $\pm$ 3 km/s. The results diverge by approximately 10\%. It must be noted that the velocity dispersion profiles generally show a very large scatter with different galaxies \citep[see][Falc\'{o}n-Barroso et al.]{Cappellari2006}. We therefore also decided to derive $\sigma_{\mathrm{e}}$ using another method.

JAM can be used to predict $V_{\mathrm{RMS}}$, V and $\sigma$ maps of $10'' \times 10''$ size, sufficiently enough to contain the effective radius of $3.9''$. We used the luminous mass model (for details see Sect.~\ref{mass model},~\ref{Jeans model}) and constrained the JAM model with the GMOS FOV.
This model reproduced the general features (increased sigma lobes in the center, decreasing velocity dispersion with increasing radius) of the GMOS velocity dispersion map quite well. The value for $\sigma_{\mathrm{e}}$ was then derived by adding up the luminosity-weighted pixel values within an aperture radius of the effective radius of the bulge of $3.9 \pm 1.4\arcsec$. This method yielded an effective stellar velocity dispersion of 113 $\pm$ 5 km/s.

Previous measurements predict that the central velocity dispersion for NGC 4414 in a rectangular aperture of size $2'' \times 4''$ yields  $\sigma_{\mathrm{c}}=117 \pm 4$ km/s  \citep{Barth2002,Ho2009}. This value is consistent with the effective stellar velocity dispersion based on the JAM model and from the power-law extrapolation for early-type galaxies. As a result of the given analysis we conclude the effective stellar velocity dispersion of NGC 4414 to have an averaged value of 115.5 $\pm$ 3 km/s.

\section{DYNAMICAL MODELING} \label{dynamical modelling}

In this Section, we present the dynamical models which were constructed to measure the mass of the central black hole. In order to assess the robustness of the results, we used two methods which contain different assumptions and therefore provide independent results. The first method is the predefined JAM method \citet[][(see footnote 2)]{Cappellari2008} which is based on the Jeans equations \citep{Jeans1922}. In the second method, we applied the more general Schwarzschild orbit superposition method \citep{Schwarzschild1979,Cretton1999,vanderMarel1998,Gebhardt2003,Cappellari2007,Onken2014} which is a complex numerical realization of the central galactic orbits. Both methods are further described in their allocated sections. A common requirement for both methods is the determination of the stellar gravitational potential which can be derived from the stellar luminosity of the galaxy combined with its M/L, which is discussed in the following Section.
\subsection{The luminous mass model} \label{mass model}
\label{ss:mge}

We modeled the dust-corrected galaxy surface brightness by using the Multi-Gaussian Expansion (MGE) introduced by \citet{Monnet1992} and \citet{Emsellem1994} . In order to simplify the convolution and deprojection calculations the projected surface brightness is parametrized as a sum of two-dimensional concentric Gaussians
\begin{equation}
\Sigma(x',y')=\sum^{N}_{j=1}\frac{L_j}{2\,\pi\sigma _{j}'^{2}\,q'_j}\exp\left[-\frac{1}{2\sigma_j'^2}\left(x'^2+\frac{y'^2}{q_j'^2}\right)\right]
,\end{equation}
where N is the number of Gaussians with each having the total luminosity $L_j$, an observed axial ratio between $0 \leq q'_j\leq 1$ and a dispersion $\sigma'_j$ along the major axis. 

We performed our MGE modeling of NGC~\,4414 by using the software and method developed for general application on galaxies by \citet[][(see footnote 2)]{Cappellari2002}. A well-constructed dynamical model requires the combination of deep imaging of large FOV data with high-resolution data in the center of the galaxy. This provides a good estimate of the M/L in the outskirts and also reveals the nuclear morphology of the galaxy. In order to properly decompose the components of the light profile, we used the MGE routine on the ground-based wide-field SDSS r-band ($12' \times 12'$) and high-resolved HST/WFPC2 F606PC ($36.4'' \times 36.4''$) image simultaneously. Due to the large amount of dust, we used the HST image only to fit the innermost isophotes ($R\leq 5''$) and the dust-corrected SDSS r-band to measure the shape of the outer-disc isophotes. For the inner and outer regions, we kept the same constraints for the axial ratio of the two-dimensional Gaussians. As the dust attenuation significantly changes the shape and the amount of measured light, we applied a dust mask to the contaminated regions (see Appendix~\ref{HSTdust}). Before the MGE model of equation (1) can be compared with the observed surface brightness, the instrumental and atmospheric PSF have to be taken into account. We obtained a model of the HST/ WFPC F606PC PSF at the center of the galaxy using the Tiny Tim HST PSF modeling tool \citep{Krist2001} and parametrized it with a circular MGE model (see Appendix \ref{HST PSF}). 

\begin{table}
\caption{MGE parameters for the deconvolved combined HST/SDSS surface brightness of NGC~\,4414: index of Gaussian $j$, surface brightness $I'_j$ , dispersion $\sigma'_j$, axial ratio $q'_j$ and total luminosity $L'_j$ given for each Gaussian }
\centering
\begin{tabular}{ccccc}
\hline\hline
$j$   & $I'_j$ & $\sigma'_j$ & $q'_j$ & $L_j$ \\
        (Pixels)& $(L_{\odot , r}pc^{-2})$     & (arcsec) & & $(\times 10^9 L_{\odot,r})$    \\
\hline
$1$  & $151393$ & $0.0565$ & $0.806$ & $0.019$ \\
$2$  &$36575$ & $0.116$  & $0.748$ &  $0.018$\\
$3$  &$47539$  & $0.245$  & $0.6$  & $0.082$ \\
$4$  &$13226$ & $0.325$ & $1.0$ & $0.067$ \\
$5$  &$10810$ & $0.75$ & $0.6$ &  $0.175$\\
$6$  &$5231$ & $1.08$ & $0.91$ & $0.266$\\
$7$  &$3212$ & $3.07$ & $0.82$ & $1.18$\\
$8$  &$658$ & $13.2$ & $0.6$ & $3.306$ \\
$9$ &$285$ & $31.9$ & $0.6$ & $8.335$ \\
$10$  &$32$ & $56.9$ & $0.6$ &  $2.982$\\
$11$ &$13$ & $76.6$ & $0.831$ &  $2.952$ \\
\hline
\end{tabular}
\label{mge2}
\end{table}
\begin{figure*}
\centering
   \includegraphics[width=\hsize]{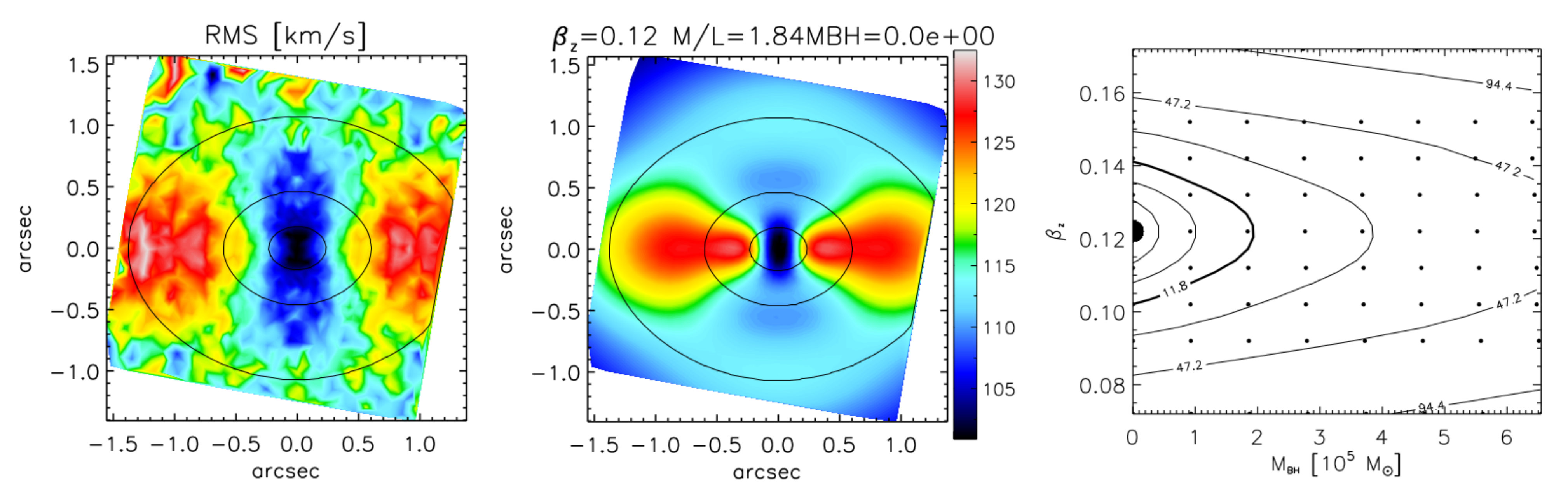}
      \caption{Results of the JAM modeling method. The left panel shows the bi-symmetrized $V_{\mathrm{rms}}=\sqrt{V^2+\sigma^2}$ observed with NIFS. The middle panel shows the JAM model for the upper limit of the mass of the central SMBH $M_{\mathrm{BH}}= 0\,M_{\odot}$ and the best fitting anisotropy parameter $\beta =0.12$. It is clearly visible that the model cannot reproduce the NIFS $V_{\mathrm{RMS}}$ very well. The right panel shows the grid of computed models (black points) overplotted with smoothed $\Delta \chi^2$ contour lines to find the best fitting model (larger black dot). The smoothing was applied by a minimum curvature algorithm. The thick contour line denotes the $3\sigma$ threshold.}
         \label{jam}
\end{figure*}
The final MGE model is composed of eleven concentric Gaussians. Except for the dust structures on the eastern side of the galaxy and the spiral arms, the MGE model reproduces the shape of the isophote contours very well (Fig.~\ref{mgefit}). The best fitting MGE parameters of NGC~\,4414 in physical units are listed in Table~\ref{mge2} following prescriptions in \citet{Cappellari2002}.

We then de-projected the derived MGE model by using the MGE formalism to retrieve the three-dimensional intrinsic galaxy luminosity density. The de-projection provides non-unique solutions \citep{Rybicki1987} from which the MGE fit then chooses a density profile which is consistent with the observed photometry.

Finally, the galaxy luminosity density can be converted to the galaxy total mass density by multiplication with the galaxy's stellar $(M/L)$, which can be different for each Gaussian. In our models, we assume a constant M/L. The luminosity density is used in the next sections to construct dynamical models of NGC~\,4414.               

\subsection{Dynamical Jeans model} \label{Jeans model}
The first method to derive the mass of the central black hole is based on the Jeans (1922) equations which follow on from the steady-state collisionless Boltzmann equation. The Jeans equation relates the galactic gravitational potential to the second velocity moment and the galaxy luminosity density \citep{Binney2008}. The additional anisotropy parameter $\beta_z$ measures the anisotropy of the velocity distribution and therefore the orbital distribution in the galaxy. The solution of the line-of-sight integral over each Gaussian component is given by equation (28) in \citet{Cappellari2008}. The second velocity moment is well approximated by the observable quantity
\begin{equation}
V_{\mathrm{rms}}=\sqrt{V^2+\sigma^2}
,\end{equation}
where V is the stellar mean velocity and $\sigma$ the velocity dispersion which parametrize the LOSVD and were measured in Section~\ref{kinematics}. \citet{Cappellari2008} provide the JAM software (see footnote 2) which allows to construct a model of the central dynamics of NGC~\,4414 based on the Jeans formalism. Although the JAM method does not allow for a general anisotropy and could, in principle, produce biased results, it was shown to provide results in agreement with more general techniques in the cases where it was compared \citep{Cappellari2010,Seth2014,Drehmer2015}. Of particular interest is the case of the black hole in NGC 1277, where the Schwarzschild's model of \citet{vandenBosch2012} indicated a significantly larger black hole than the models by \citet{Emsellem2013}, based on an N-body realization having the same 1st and 2nd moments of a JAM model (computed via Eqs. 19 to 21 in \citet{Cappellari2008}), which generalizes the MGE formalism. The latter black hole estimate turned out to be confirmed by subsequent work \citep{Walsh2016} using high-resolution IFU data. This illustrates the usefulness of comparing black hole determination using different techniques based on different assumptions.

\begin{figure*}[!htb]
\centering
   \includegraphics[width=\hsize]{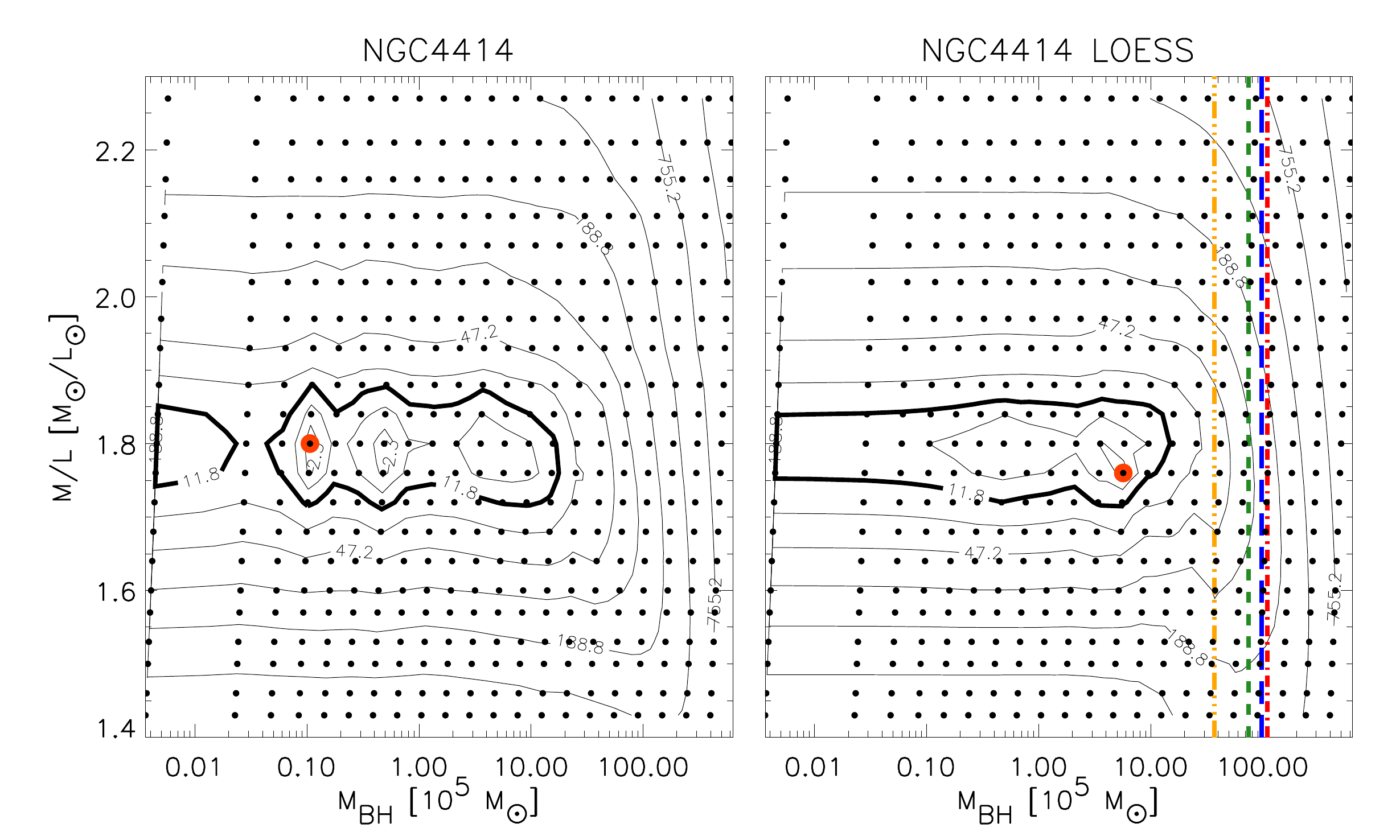}
      \caption{Grids of Schwarzschild dynamical models (round symbols) with different mass-to-light ratios and black hole masses. Contours are the $\Delta\chi^2 = \chi^2 - \chi^2_{\rm min}$ levels. The thick contours are the $3\sigma$ levels. Both panels are for regularization parameter of $\Delta=10$, while the right hand panel shows smoothed  $\Delta\chi^2$ contours using the local non-parametric regression LOESS algorithm \citep{Cleveland1979,Cappellari2013}, with the local polynomial set to quadratic. The large red circle shows the model with the formal $\chi^2$ minimum. The LOESS smoothed $\chi^2$ contours indicate that no $\chi^2$ minimum is reached within our grid when models are marginalized over all M/L ratios. The dashed (green) line indicates the values of M$_{\rm BH}$  for which the sphere of influence is three times smaller than the resolution of our NIFS data (this value is also close to the prediction based on the \citet{McConnell2013} M$_{\rm BH} - \sigma_{\mathrm{e}}$ scaling relation). The dashed-double dotted (orange), long-dashed (blue) and dashed-dotted (red) vertical lines indicate predictions for M$_{\rm BH}$ of the \citet{VandenBosch2016}, \citet{Saglia2016} and \citet{Greene2016} scaling relation (for all galaxies in their samples), respectively.}
         \label{f:grid}
\end{figure*}
\begin{figure*}[!htb]
\centering
   \includegraphics[width=\hsize]{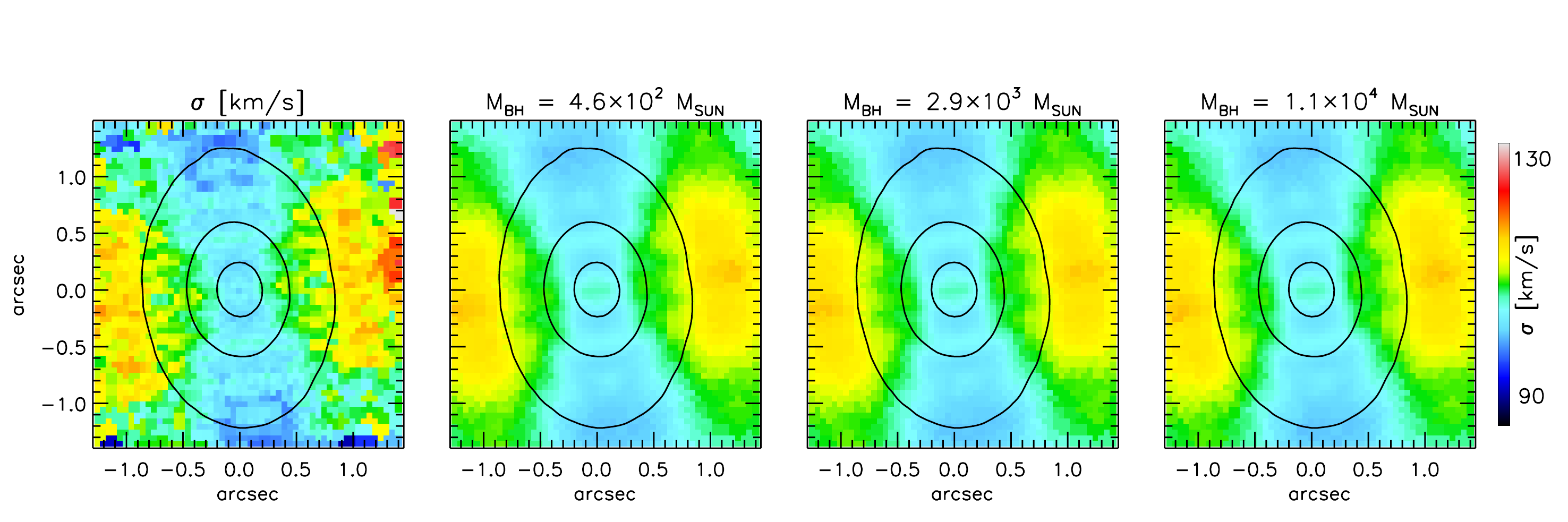}
      \caption{Comparison between the observed NIFS velocity dispersion map and three low M$_{\rm BH}$ Schwarzschild models, all at the same M/L=1.80. From left to right: the data, the model with the lowest M$_{\rm BH}$ used in this work, a model which is formally outside the three-sigma uncertainty level, and the formal best fit model from Fig.~\ref{f:grid} left panel. The M$_{\rm BH}$ values are given above the maps. There are no significant differences between these three models, and the differences with the data are on the level of systematics, arguing that the the three-sigma uncertainty level on the left hand panel of Fig~\ref{f:grid} should be continuous and no lower limit to the M$_{\rm BH}$ is found in NGC\,4414.  }
         \label{f:lowmass}
\end{figure*}
From the derived MGE surface brightness we constructed several axisymmetric models. The models have three free parameters which are 1) the anisotropy parameter $\beta_z$, 2) the mass of the black hole $M_{BH}$ and 3) the dynamical mass-to-light ratio M/L. We fixed the inclination of the galaxy to that of the large-scale atomic and molecular gas disk \citep[][Sect.\ref{discussion} of this paper]{Vallejo2002,Wong2004}.

In total, we constructed 54 dynamical models in a grid of $\beta_z =[0.07, 0.17]$ and $M_{\mathrm{BH}}=[0,3\times 10^5\,M_{\odot}]$. The formally best fitting model parameters could be determined by minimizing $\chi^2$ between model $V_{\mathrm{rms,model}}$ and the NIFS observation $V_{\mathrm{rms,obs}}$.
Figure~\ref{jam} illustrates the $\beta_z-M_{\mathrm{BH}}$ grid of our constructed models. The JAM models are plotted as black points. In order to describe the agreement between the JAM models and the observed $V_{\mathrm{rms}}$, we overplotted the contours of $\Delta \chi^2=\chi^2-\chi^2_{\mathrm{min}}$ on the model grid with $\chi^2_{\mathrm{min}}$ defined by the best fitting model. The mass of the black hole is not constrained resulting in an upper limit measurement of $M_{\mathrm{BH}}=1.5 \times 10^5 M_{\odot}$ (for one degree of freedom, marginalizing over $\beta_z$) and $M_{\mathrm{BH}}=2 \times 10^5 M_{\odot}$ (for two degrees of freedom, see Fig.~\ref{jam},) at 3 $\sigma$ significance obtained solely from the NIFS observations and the JAM models. This is approximately 50 times lower than that predicted by the $M_{BH}-\sigma_{\mathrm{e}}$ relation \citep[e.g.][late-type]{Greene2016}. In addition, JAM provides the dynamical M/L of the best-fitting model of $M/L=1.84 \pm 0.04$.

For a visual comparison, we show the $V_{\mathrm{rms}}$ data and the best fitting JAM model in Fig.~\ref{jam}. Formally the best fitting dynamical model is given by $M_{\mathrm{BH}} = 0\, M_{\odot}$ and $\beta =0.12$. We note that the JAM models generally do not reproduce the observations very accurately, but resemble the observed $V_{\mathrm{RMS}}$. However, while the actual shape of the $V_{\mathrm{rms}}$ cannot qualitatively be reproduced by the JAM models, the distinct $V_{\mathrm{rms}}$ drop from the center of the observations is also present in the models. This drop is very crucial as it provides important implications on the central black hole. Models with higher black hole masses fail more distinctly to reproduce this drop in the $V_{\mathrm{rms}}$ (see Sect.\ref{resolution limit} and Fig.~\ref{f:bh}).

\subsection{Schwarzschild models} \label{Schwarzschild model}

We constructed dynamical models based on the \citet{Schwarzschild1979} method, generalized to fit stellar kinematics \citep{Richstone1988,Rix1997,vanderMarel1998}. The implementation we use was optimized for integral-field observations and is described in \citet{Cappellari2006}. Briefly, the first step of the method is a construction of a library of orbits which evenly sample the space of three integrals of motion: energy $E$,  vertical projection of the angular momentum $L_{z}$ and the third integral, $I_{3}$. Energy is sampled at 41 logarithmically spaced points specifying the representative radius of the orbit, while at each energy we used eleven radial and eleven angular points for sampling $L_z$ and $I_3$, respectively. Each model library comprises a total of 2143150 prograde and retrograde orbits bundled in groups of $6^3=216$ orbits with adjacent initial conditions. The orbits are integrated in the potential defined by the MGE parametrisation of the light distribution (Section~\ref{mass model}), which was projected at an inclination of 55 degrees, while assuming axisymmetry. The two free parameters which define the total potential are the mass-to-light ratio (M/L) and the mass of the black hole. Once the orbits are integrated, they can be projected onto the observable space (position in the sky and the LOSVD parameters), while taking into account the PSF and the Voronoi bins. Each orbit is assigned a weight in a non-negative least-squared fit \citep{Lawson1974}, and when combined they reproduce the observed stellar density and kinematics in each bin. Both NIFS and GMOS kinematic data are used to constrain the models, however we exclude the GMOS data from the central 0.8 arcscec (we verified that including the GMOS data does not change the results). As our Schwarzschild model is axisymmetric by construction, we first symmetrized the kinematics ($V$, $\sigma$, $h_3$ and $h_4$ maps) using point-(anti)symmetry around the photometric major axis, averaging at positions: [($x$,$y$),($x$,$-y$),($-x$,$y$), ($-x$,$-y$)], while keeping the original errors in each bin. Finally, when running the linear orbital superposition, we used two levels of regularization, a moderately low $\Delta=10$ and a high one $\Delta=4$ (as defined in \citet{vanderMarel1998}). Both regularizations gave equivalent results and we present the one with moderate regularization.

Grids of 441 dynamical models are presented in Fig.~\ref{f:grid}. The contours show the $\Delta\chi^2$ levels, which are calculated for a two-parameter distribution, while the thick contour displays the $3\sigma$ level. In the left hand panel there is a discontinuity of the $3\sigma$ $\Delta\chi^2$ level at lower masses, suggesting the possibility of also constraining the lower limit of M$_{BH}$. However, this is not real and likely a spurious result of the models. Running a smaller orbital library constructed by sampling the three integrals ($E$, $L_z$, $I_3$) with (21,7,8) starting conditions, results in several of these local minima in the region below M$_{BH} < 10^6$ M$_\odot$. Regularizing the Schwarzschild models does not significantly improve the grid of $\Delta\chi^2$ values. The grid of $\Delta\chi^2$ becomes, however, significantly smoother when the size of the orbital library is increased, as shown here. A further increase of the library size would likely remove the present discontinuity. 
\begin{figure*}[!htb]
\centering
   \includegraphics[width=1.\hsize]{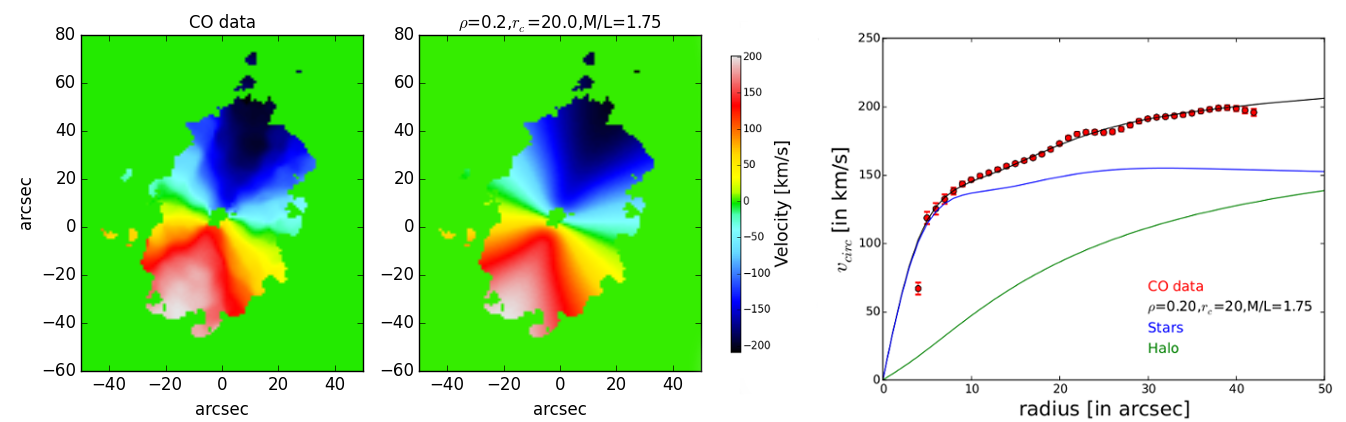}
      \caption{Comparison between the CO observations and our convolved two-dimensional velocity model. The left and middle panel show the CO data  obtained by \citet{Wong2004} and our model which best describes the CO data. The right panel shows the rotation curves of the CO observations (red points) and the best-fitting model (black solid line). We also separately plot the rotation curves from the stellar (blue solid line) and dark matter (green solid line) potential. All model curves are convolved with the kernel by \citet{Qian1995}.}
         \label{dm}
\end{figure*}
As the increase in the orbital library is expensive in terms of computing, we examined the difference between these models in Fig~\ref{f:lowmass}. Next to the observed velocity dispersion (NIFS), we plot two models with the lowest M$_{BH}$ and the model of the formal best fit from the left hand panel on Fig~\ref{f:grid}. It is obvious that these models are not different in any significant way, even though the formal $\chi^2$ value is sufficiently higher for the model in the middle to exclude it from the three-sigma uncertainty level. The differences between the models are at the level of the systematic errors affecting the data, and this is likely connected to the size of the orbital library and intrisinc degeneracies such as the deprojection at a relatively low inclination (see \citet[][]{Krajnovic2005} for a similar situation related to the degeneracy in recovering the inclination and \citet{vanderMarel1998} for a discussion on the topology of $\Delta\chi^2$ contours due to finite numerical accuracy of the models).  
 
Fig.~\ref{f:lowmass} suggests that one could apply a moderate level of smoothing to improve the topology of  $\Delta\chi^2$ contours \citep[e.g.][]{Gebhardt2003}. For the right hand plot of Fig.~\ref{f:grid} we adaptively averaged $\Delta\chi^2$ contours using the local regression smoothing algorithm LOESS \citet{Cleveland1979}, adapted for two dimensions \citep{Cleveland1988} as implemented by \citet{Cappellari2013b} (see footnote 3). We used a limited number of points for the LOESS kernel with the fraction (of total number of points) equal to 0.2, and a local quadratic approximation, as the location of the $\chi^2$ minimum is similar to a quadratic function. This produces a smooth $\Delta\chi^2$ with a well-defined upper limit, fully consistent with the original grid. The formal best fit now has a somewhat larger $M_{\rm BH}$, and a slightly smaller M/L, but these should not be taken literally as there is no sufficient difference between the models within the three-sigma uncertainty level (see also the first model on Fig.~\ref{f:bh}). 

The Schwarzschild models constrain the mass-to-light ratio to M/L $= 1.80\pm0.09$, in good agreement with other estimates (Section~\ref{errorbudget} and Table~\ref{ml_table}). As discussed, we do not constrain the lower mass limit of $M_{\rm BH}$ at $3\sigma$ level, while the upper mass limit is $1.56\times10^{6}$ M$_{\odot}$. In  Appendix~\ref{app:models} we show the comparison between a representative Schwarzschild model within the three-sigma uncertainity level having M/L $= 1.80$ and M$_{BH} = 5.81\times10^5$ M$_{\odot}$, and the kinematics from GMOS and NIFS observations.

\section{DISCUSSION} \label{discussion}
Both the Jeans and the Schwarzschild models from our previous analysis provide consistent results for the black hole mass of NGC~\,4414, constraining only the upper limit value. These upper limits mark the mass boundary of where the models start to become clearly inconsistent with our data. Below we carry out a comprehensive error analysis of our measurements, discuss our results with respect to the resolution limit and place the galaxy on the most recent and relevant $M_{\mathrm{BH}}$-host galaxy relations.

\subsection{Error budget}\label{errorbudget}
Many assumptions and uncertainties go into the dynamical models to determine the mass of the central black hole. While the statistical errors given in the previous sections account for the uncertainties of the models, it is also important to take a closer look at systematic effects which can drastically change the results. In this section, we evaluate the importance and effects of dust contamination, distance accuracy, dark matter contribution, variation in M/L with radius, and kinematical tracers.\\
\textit{Dust Contamination:} A large amount of dust pollutes the facing side of NGC 4414 (see Fig.~\ref{dustcorrection2}, Fig.~\ref{dustcorrection3}). This especially affects the visual light WFPC2 F606W PC and SDSS r-band images, less so in the infrared. Therefore, we corrected both images before modeling the surface brightness (see Appendix~\ref{HSTdust}). We checked the effect of the dust-correction by also creating MGE models from the WFPC2 F606W image in combination with different dust masks and even without a dust mask. Constructing dynamical models for these modified MGE models, however, revealed no significant change in the upper limit measurement, but the models represented the data less well.\\
\textit{Distance:} NGC 4414 has approximately 40 distance measurements based on Cepheids and Tully-Fisher methods which span a range between 5 and 25 Mpc. Taking only distances into account that have a conservative error below 1 Mpc, the distance span lowers to between $16.6\pm 0.3 $ and $21.1\pm 0.9$ Mpc \citep{Kanbur2003,Paturel2002}. This gives a difference of [-1.7,+4.0] Mpc to our applied distance value of 18.0 Mpc. In our dyncamical models, the distance operates as scaling factor and is directly proportional to the mass of the black hole and anti-proportional to the M/L. This means the uncertainty from the distance is around 20\%. Taking the distance uncertainty into account for our models, we get $M_{\mathrm{BH}}\le 1.85\times10^6$ M$_{\odot}$ and $M/L=1.8\pm 0.35$.\\
\textit{Dark Matter}:
Both dynamical modeling methods do not explicitly take dark matter (DM) into account. Consequently, the models are only reliable when the shape of the total density distribution is well approximated by the stellar density distribution within the region where we fit the kinematics.
However, different studies \citep{Gebhardt2009,Schulze2011a,Rusli2013} report a change in black hole mass measurements when including the presence of dark matter in the region covered by the kinematic data. This happens when the dynamical models include kinematics at sufficiently large radii that the difference between the slope of the total and stellar density becomes significant. The reported effect of accounting for dark matter in the dynamical modeling method is a factor between 1.2 and 2 in the black hole mass and results from the degeneracy between the dark matter halo mass, the stellar mass-to-light ratio and the black hole mass. Therefore, neglecting the dark matter component increases the M/L leading to a smaller $M_{\mathrm{BH}}$ to fit the observed kinematics. \citet{Vallejo2003} fit a Navarro–Frenk–White halo \citep{Navarro1996} to the rotation curve of NGC~\,4414 and report a low-mass DM halo in the central region of the galaxy. As the low-DM assumption is important for the black hole measurement, we further validated our dynamical results by deriving the dark matter fraction within the observed region of NGC~\,4414.

Therefore, we constructed a model circular velocity map and compared it to CO observations deriving the stellar mass-to-light ratio and probing the existence of dark matter in the center of the galaxy. The CO velocity fields were derived by \citet{Wong2004} from mm-interferometric observations, which cover the CO emission for a large portion of the visible galaxy ($R \approx 50''$). By fitting Gaussians to the spectrum of each pixel, using a customized version of the MIRIAD task GAUFIT, the authors derived the velocity map of their data. We used the mass model from Section~\ref{mass model} to derive the circular velocity profile along the galaxy major axis $v_{\mathrm{mj}}$. The central DM contribution was parametrized with a pseudo-isothermal sphere which predicts the following circular velocity for the DM component:
\begin{equation}
V_{\mathrm{c,DM}}^2(r)=4\pi G\rho_0 r_c^2 \left[1-\frac{r_c}{r} \arctan(\frac{r}{r_c}) \right]
,\end{equation}
where $\rho_0$ denotes central mass density of the sphere and $r_c$ is the core radius.  We added the contribution of the pseudo-isothermal sphere quadratically to the stellar circular velocity.
Using the standard projection formula 
\begin{equation}
v_{\mathrm{c,stars}}(x,y)=v_{\phi}\left(\frac{x \sin i}{r}\right)=v_{mj}\left(\frac{x}{r}\right)
,\end{equation}
with the radius $r^2=x^2+(y/\cos i)^2$ and galaxy inclination $i$, it is possible to construct an axisymmetric two-dimensional velocity map model from the velocity profile \citep{Krajnovic2005}. The \citet{Wong2004} CO velocity field is sampled on square pixels of 1 arcsec size, while the observations had a beam of $6.53 \times 4.88$ arcsec. We convolved our model velocity field with the circular beam size of 6.53 arcsec using the kernel of \citet{Qian1995}. The constructed velocity field is a function of stellar M/L, galaxy inclination $i$, central DM density $\rho_0$ and core radius $r_c$. In order to find the velocity field model which best represents the CO data, we tested different parameter grids and finally obtained $M/L=1.75$, $i= 55^{\circ}$, $\rho_0=0.2$ and $r_c=20 ''$. The degeneracy between stellar mass-to-light ratio and dark matter was clearly visible while constructing the different models. The derived value for the M/L is consistent with the other measurements in this paper. In Figure~\ref{dm}, we present the CO data, the best matching two-dimensional velocity field and a cross section of our velocity field model (black line) and the CO data (red dots) along the major axis. Except for the most central point of the CO data, our model reproduces the entire CO data very well. However, this outlier is tightly related to the CO emission hole in the center of NGC~\,4414. Figure~\ref{dm} also illustrates the stellar and dark matter contribution of the velocity profile separately. We note that, for $r<10''$, which is probed by the NIFS and GMOS data, the dark matter contribution is small; approximately 10\%\ of the total mass. 

We also tested the DM  content of NGC 4414 by running another set of dynamical JAM models, this time adding a spherical dark halo component, modeled as generalized Navarro–Frenk–White profile \citep{Navarro1996} with fixed $r_s=20$ Kpc, to the galaxy potential. We then calculated a parameter grid of $M_{BH}$, M/L, the halo density ${{\rho }_{s}}$ at $r_s$ and the dark halo slope $\gamma$ for $r\ll {{r}_{s}}$. 
These models showed consistent results with the CO data.
Therefore, the no-dark-matter assumption is acceptable for the dynamical models of the central region of NGC~\,4414. This is consistent with other findings \citep{Vallejo2003,Cappellari2013b}.

\textit{Stellar M/L Variation:} Our dynamical models assume that the M/L stays constant for different radii. However, stellar population changes can result in different M/L. Various observations and models suggest that different evolutionary histories in different regions of spiral galaxies lead to variations in the stellar M/L \citep{Bell2001}, which can affect the results of the dynamical models \citep{Portinari2010}. We tested the M/L variation by applying pPXF to our GMOS data and fitting a linear combination of Simple Stellar Population (SSP) model spectra to the galaxy spectrum.

\begin{figure}
\centering
   \includegraphics[width=\hsize]{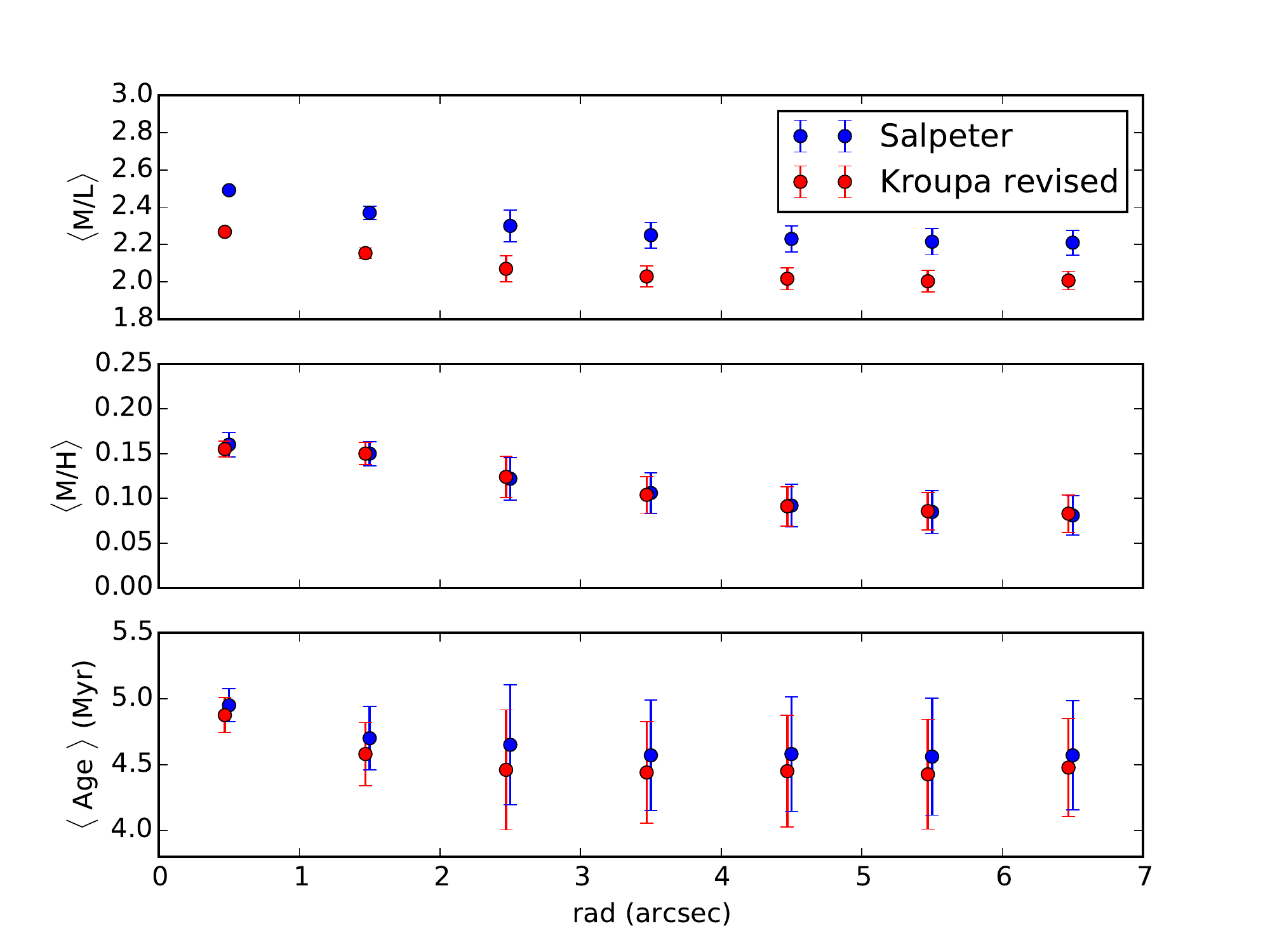}
      \caption{Radial trends of the stellar M/L, the mass-weighted metallicity and the mass-weighted age of NGC 4414 derived from pPXF. The given radius is the radius of a circular aperture around the galaxy center which defines the spectra used for the respective composite spectrum. We tested the Salpeter IMF (blue) as well as the Kroupa revised IMF (red). For improved clarity, the red points have been shifted very slightly to the left.}
         \label{radialtrend}
\end{figure}
\begin{table}
\caption{Summary of the derived M/L in NGC 4414}
\centering
\begin{tabular}{lcc}
\hline\hline
Method &  M/L  &  Radius [''] \\
\hline
JAM + NIFS &  1.84  $\pm$ 0.04  & 1.5\\
JAM + GMOS & 2.14    $\pm$ 0.04 & 5.5 \\
Schwarzschild & 1.8  $\pm$ 0.09 & 5.5\\
Stellar populations & 2.15  $\pm$ 0.03 & 1.5\\
Stellar populations  & 2.0  $\pm$ 0.06 & 5.5\\
CO map + DM & 1.75  $\pm$ 0.02 & 40\\
\hline
\end{tabular}
\tablefoot{Due to the distance uncertainty, the M/L uncertainties increase by approximately 10\%.} 
\label{ml_table}
\end{table}

\begin{figure*}[!htb]
\centering
   \includegraphics[width=\hsize]{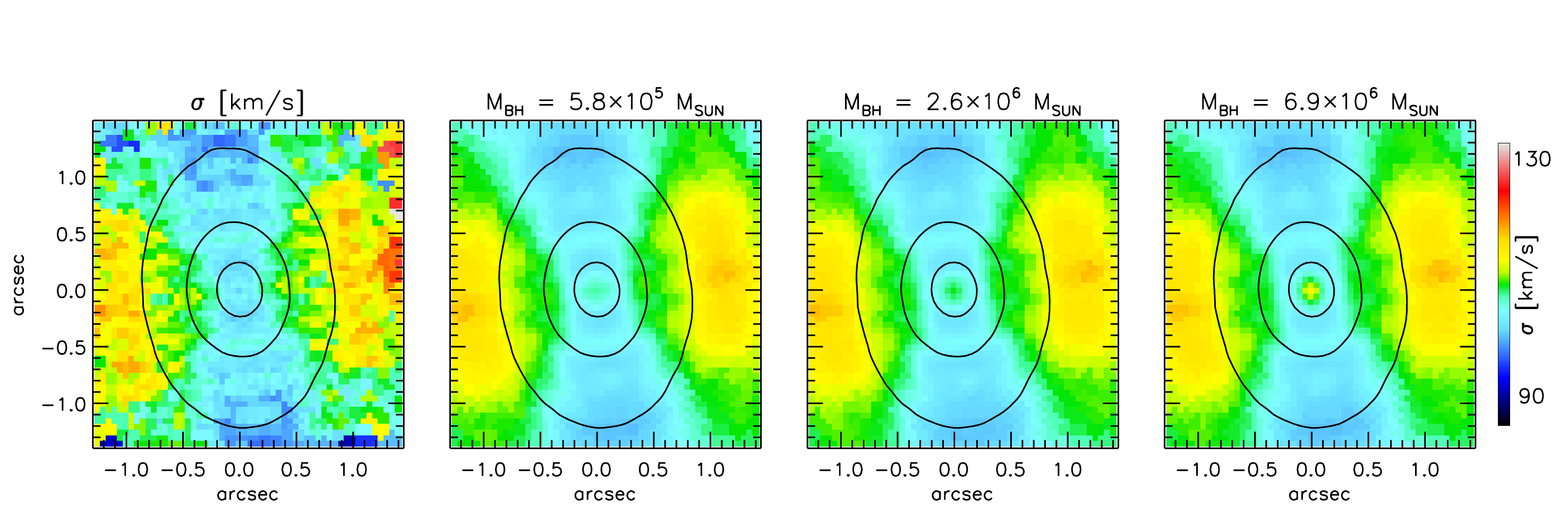}
      \includegraphics[width=\hsize]{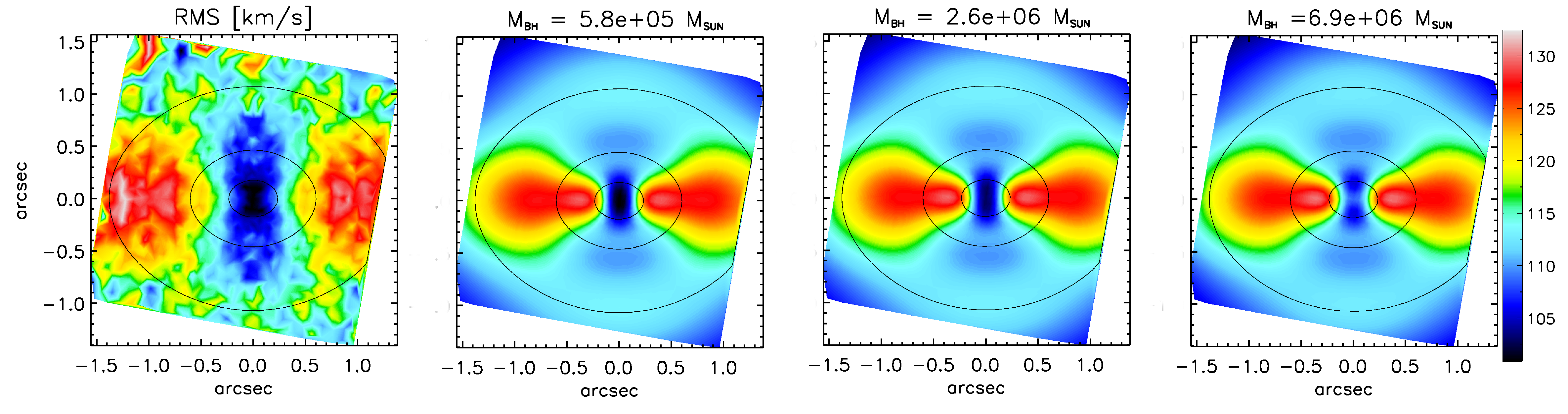}
      \caption{Comparison between the observed NIFS velocity dispersion map and three representative Schwarzschild models, all at the same M/L=1.80 and with different black hole masses. From left to right: the data, the formal best fit M$_{\rm BH}$ (using LOESS) model, a model with M$_{\rm BH}$ a factor of five higher than the previous one (70\% higher than the derived upper limit), and the model with M$_{\rm BH}$ for which the sphere of influence is the same as our spatial resolution. The M$_{\rm BH}$ values are given above the maps. Note the appearance of the central velocity dispersion peak for models above the formal upper limit. The same comparison is given for the JAM $V_{rms}$ models in the lower panels. }
         \label{f:bh}
\end{figure*}

 In order to cover the optical wavelength range of the GMOS spectrum, we used the model spectra from the MILES SSP model library \citep{Vazdekis2010} which spans an equally spaced $50 \times 7$ grid of age ranging from 0.06 to 17.78 Gyrs and metallicities between [Z/H]= -2.32 and 0.22. As initial mass function (IMF) we assumed the standard \citet{Salpeter1955} IMF and the \citet{Kroupa2001} revised IMF. We used the method described in \citet{McDermid2015,Shetty2015} to determine the stellar population of NGC~\,4414, by applying weights to the different SSP model spectra and constrain these with the pPXF built-in regularization option which specifies the penalization of the $\chi^2$. Models with similar ages and metallicities were assigned smoothly varying weights until difference in $\chi^2$ between the current and non-regularized solution satisfied the following criteria $\Delta \chi^2 \approx \sqrt{2 N,}$ where N is the number of pixels fitted in the spectrum \citep{Press2007}.\\
We analyzed the stellar populations using two different approaches. In the first approach, we fitted the SSP models to the composite spectrum of the entire GMOS cube from which we determined the mass-weighted age of NGC 4414 to be approximately 4.5 Gyrs and the mass-weighted metallicity of $\langle[M/H]\rangle = 0.0975$ for Salpeter as well as 4.5 Gyrs and  $\langle[M/H]\rangle = 0.0812$ for Kroupa. The derived stellar population provides the mass-weighted M/L, calculated as
\begin{equation}
(M/L)= \frac{\sum_j w_j M_{*,j}}{\sum_j w_j L_{r,j}}\quad,\end{equation}
where $w_j$ denotes the weight given by the pPXF fit to the $j$th template, $M_{*,j}$ is the mass of the $j$th  template given in stars and stellar remnants, and $L_{r,j}$ is the r-band (AB) luminosity of the template. We determined a M/L = 2.1 from the stellar population. 
In our second approach we were interested in the strength of the M/L gradient for NGC 4414. Therefore, we co-added the spectra in circular apertures of different radii and applied pPXF to these composite spectra. Figure~\ref{radialtrend} shows the radial trends for the derived stellar M/L, the mass-weighted metallicity and the mass-weighted age for the two different assumed IMFs.The difference in IMF solely introduces an offset between the results. Clear evidence for a gradient in M/L is visible in Fig.~\ref{radialtrend}, however, the stellar M/L only changes by 10\%\ between 0.5\arcsec and 6.5\arcsec. The uncertainty in the black hole mass due to the stellar M/L variation also yields approximately 10\%. Therefore, we conclude that within the level of precision achievable with our dynamical modeling methods, M/L=const. is an adequate assumption. By keeping the dynamical M/L constant, we are also able to account for the small amount of dark matter found in the center of the galaxy \citep{McConnell2013b}. 

Table \ref{radialtrend} gives an overview of the M/L measurements of NGC 4414 which we determined with very different methods. The measurements show great consistency with each other. However, the stellar M/L values derived from the stellar populations show noticeably higher trends than the dynamical M/L. The dynamical M/L measurements are strongly dependent on the distance accuracy of NGC 4414. As noted above, this distance uncertainty increases the measured dynamical M/L uncertainty by 10\%. Thus, assuming a distance of 16 Mpc instead would increase the Schwarzschild M/L to $2.0 \pm 0.1$ which is more consistent with the stellar M/L. Therefore, we conclude that the different M/L ratios result from uncertainties in the different methods.

\textit{Kinematic Tracer:}
In order to construct dynamical models, we would ideally use a luminosity density which was observed in a similar band to the kinematical tracer. Using a luminosity density in the r-band while probing the central kinematics in the K-band could lead to tracking different stellar populations. 
Therefore, we tested the compatibility between the r-band and K-band surface brightness of NGC 4414. Figure~\ref{PSF} provides initial evidence of how well the the HST data matches the NIFS K-band image. After convolution with the NIFS PSF, the HST data fits the center of the collapsed NIFS data very well and discrepancies arise at the edges of the NIFS data where dust begins to emerge. 
In a second attempt to compare both tracers, we constructed a MGE model from the combination of the reconstructed NIFS image, the WFPC2 F606W image and the SDSS r-band image. We used the reconstructed NIFS image to probe the central 1.2\arcsec of NGC~4414, HST between $1.2\arcsec<r<5\arcsec$ and SDSS for the remaining extent of the galaxy. We then constructed dynamical Jeans models from this new MGE in combination with NIFS kinematics. From this test, we determined the best-fitting model parameters to be $M_{BH}= 0 M_{\odot}$ and M/L = 1.79. Within $3\sigma$ statistical uncertainty the $\chi^2$criterion provides an upper limit of $2.4 \times 10^5 M_{\odot}$, which is consistent with our r-band image-only MGE model results. Therefore, inaccuracies in the tracer distribution do not affect our final results.

\subsection{Bridging the resolution limit} \label{resolution limit}
The case of NGC 4414 provides important conclusions relating to the resolution limit of black hole measurements. In this study, we are able to constrain the black hole upper mass limit of NGC 4414 to a significantly lower value than the resolution limit given by our high-resolution NIFS data.

The vertical lines in Fig.~\ref{f:grid} show expected values for the M$_{\rm BH}$ in NGC~\,4414, based on the predictions from recent M$_{\rm BH} - \sigma_{\mathrm{e}}$ scaling relations \citep{McConnell2013,Saglia2016,Greene2016} and the M$_{BH}$-size-mass relation proposed by \citet{VandenBosch2016}. The closest to the derived upper limit is the M$_{BH}$-size-mass relation by \citet{VandenBosch2016}, which predicts M$_{\rm BH} = 3.66\times 10^{6}$ M$_\odot$ based on the galaxy effective radius \citep{Davis2012} and the total galaxy mass (which we obtained from JAM) . This prediction is followed by the \citet{McConnell2013} relation for all galaxy types, which, assuming the velocity dispersion of NGC~\,4414 within the effective radius, predicts M$_{\rm BH}$ = 9.4$\times 10^{6}$ M$_\odot$, approximately five times larger than our formal upper limit. A similar mass is obtained if one looks for the lower M$_{\rm BH}$ limit that we should have been able to detect, assuming that the black hole sphere of influence is three times smaller than the resolution of NIFS data (0.13\arcsec). \citet{Krajnovic2009a} showed that it is not necessary to resolve the sphere of influence to estimate M$_{\rm BH}$ if one has IFU data covering approximately one effective radius of the galaxy as well as the high resolution adaptive optics assisted IFU observations of the nucleus. With such data it is feasible to estimate M$_{\rm BH}$ for black holes when the sphere of influence is up to three times that of the spatial resolution of the data \citep[see also][]{Cappellari2010}. For our observations of NGC 4414, this lower  M$_{\rm BH}$ limit is $7.5\times10^{6} M_\odot$. Given these limits, the quality of the data presented here, and predictions from the latest estimates of the scaling relations for late-type galaxies \citep[e.g., $9.8\times 10^6$ M$_\odot$;][late-type]{Greene2016} or low mass galaxies, we would have been able to measure $M_{\rm BH}$, if it were as predicted by the $M_{\rm BH} -\sigma_{e}$ relation. However, the upper limit obtained from Schwarzschild models is still several factors smaller than the expected M$_{\rm BH}$, which suggests that the NGC\,4414 has an under-massive central black hole, if any at all. 

Furthermore, Fig~\ref{f:bh} shows that the models start significantly departing (such that it is visually easy to see the differences) from the observed data for M$_{\rm BH}$ a few times larger than the formal upper limit. The velocity dispersion map of the highest mass black hole in this figure is obviously very different from the observed velocity dispersion. The central peak is completely absent in the data. This peak decreases with smaller M$_{\rm BH}$, and is not visible for the models within the $3\sigma$ confidence level. The kinematically cold nuclear structure essentially limits the mass of the central black hole to a few times $10^6$ M$_{\odot}$, a factor of ten smaller than expected from the latest scaling relations \cite[e.g.,][]{Saglia2016,Greene2016}. The same results can be recovered from less general JAM models as well (constrained using only the high resolution data). 

Crucially, this suggests that when one uses high quality IFU data, the predicted sphere of influence should be taken only as a very rough guide for the possible M$_{\rm BH}$ one could measure. In the case of NGC\,4414, the sphere of influence for the black hole of $10^{6}$ M$_\odot$ is approximately 0.004 arcsec, while the resolution of the NIFS data is approximately 0.1 arcsec, a factor of 25 lower. Nevertheless, in this case, the models are able to rule out M$_{\rm BH} \ge 2\times 10^{6}$ M$_\odot$, as visually confirmed on Fig.~\ref{f:bh}. 

The measured upper limit for the black hole mass is approximately five times lower than the black hole we would formally be able to measure based on the sphere of influence criterion only. Constraining dynamical models with high quality IFU data and imaging (high-resolution and having a large field-of-view in both cases), can therefore bring down the measurement of black holes for at least this factor compared to the prediction based on the formal resolution limit and the sphere of influence argument. This implies that not being able to resolve the sphere of influence is not a hard limit which prohibits the detection of the corresponding massive black hole in future observations. The sphere of influence should be taken only as a rough indicator for black hole mass measurements. High Strehl ratio IFU data (covering a large area of the galaxy) can be, however, trusted to provide constraints for SMBHs several factors lower than predicted by the simple sphere of influence argument.

\subsection{Black hole - host galaxy scaling relations} \label{scaling relations}
 \begin{figure}
\centering
   \includegraphics[width=\hsize]{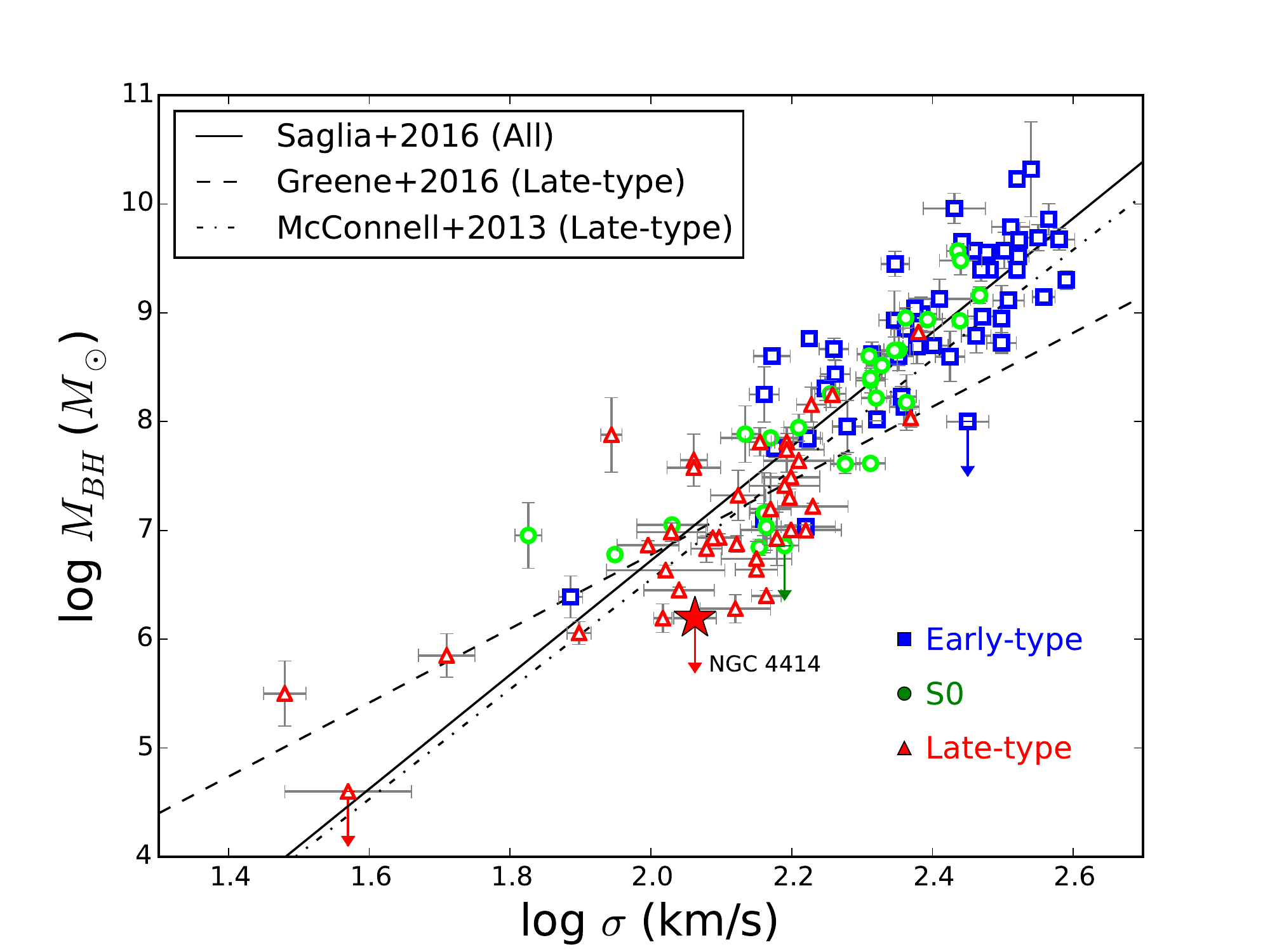}
      \caption{Location of the upper limit mass measurement of NGC~\,4414 in the $M_{BH}-\sigma_{\mathrm{e}}$ relation. The data points were taken from the BH compilation of \citet{Saglia2016} in addition to measurements by \citet{VerdoesKleijn2002,Greene2010,Coccato2006,DeLorenzi2013,denBrok2015,Walsh2015,Walsh2016,Greene2016,Thomas2016,Bentz2016} and divided into early-type galaxies (blue), S0 type (green) and late-type galaxies (red). Our upper-limit measurement is indicated by a red star. We have also added the scaling relations by \citet{Saglia2016} for all types of galaxies (solid) and \citet{Greene2016,McConnell2013} for late-type galaxies (dashed,dashed-dotted).  }
         \label{mbhsigma}
\end{figure}
The $M_{\mathrm{BH}}-\sigma_{\mathrm{e}}$ relation \citep{Ferrarese2000, Gebhardt2000, Tremaine2002} can be described by a single power law over a wide range in effective stellar velocity dispersion \citep{Graham2011,McConnell2013,kormendy2013}. Its effective stellar velocity dispersion of $\sigma_{\mathrm{e}}=115.5 \pm 3$ km/s (Sect.~\ref{bulge dispersion}) indicates that NGC 4414 contributes to the low-mass end of the $M_{\mathrm{BH}}-\sigma_{\mathrm{e}}$ correlation, which is poorly populated. Figure~\ref{mbhsigma} shows a reconstruction of the most recent version of the $M_{\mathrm{BH}}-\sigma_{\mathrm{e}}$ relation \citep{McConnell2013,Saglia2016,Greene2016} for spirals and all galaxies. We created the diagram with the SMBH sample from the compilation of \citet{Saglia2016} and added a number of spiral galaxies and previous outlier upper-limit measurements \citep{VerdoesKleijn2002,Greene2010,Coccato2006,DeLorenzi2013,denBrok2015,Walsh2015,Walsh2016,Greene2016,Thomas2016,Bentz2016}. The location of NGC~\,4414 in the $M_{\mathrm{BH}}-\sigma_{\mathrm{e}}$ relation is indicated by a red star. It is clearly visible that our $M_{BH}$ measurement lies below the shown scaling relations, but is possibly located within the region of late-type galaxies. Comparing our upper limit result with the SMBH masses derived from the scaling relations, we find a deviation of $1 \sigma$ for \citet{Saglia2016}, $1.5\sigma$  for \citet{McConnell2013} and $2.5\,\sigma$ for \citet{Greene2016} whereas \citet{Greene2016} assumes much smaller errors. Therefore, NGC 4414 could be located in the region spanned by the scatter of the relations, if not even lower. This scatter is generally very wide in the $\sigma_{\mathrm{e}}$ scaling relations, both in the low and also the upper end.

NGC~\,4414 is not the only galaxy, which shows a significant divergence from the black hole scaling relations. Many outliers can be found in the upper-mass end of the relations, but a number of galaxies have also been found for which the dynamical models predict an upper limit to the black hole mass, putting them below the $M_{BH}-\sigma_{\mathrm{e}}$ scaling relation \citep{Sarzi2001,Merritt2001,Gebhardt2001,Valluri2005,Coccato2006}. Below the scaling relation there are three upper limits of which only NGC 4414 is obtained from stellar kinematics. \citet{Vittorini2005} suggest these objects could likely be `laggard' galaxies \citep[e.g.,][]{Vittorini2005,Coccato2006} that could not yet completely develop their massive black holes. These galaxies have spent most of their lifetime isolated in `the field', where encounters are so rare that gas fueling of the galactic center is slowed down causing a limited growth of the black hole mass.

NGC 4414 seems to continue the trend seen in the late-type galaxies of \citet{Greene2016} and more work in this $\sigma$ range is necessary to better constrain the slope of the black hole scaling relations.

\section{CONCLUSIONS} \label{conclusions}
We obtained high-resolution NGS adaptive optics-assisted Gemini NIFS and large-scale Gemini GMOS IFU observations of the spiral galaxy NGC 4414 to map its kinematics which trace the gravitational potential of the central SMBH. The stellar kinematic maps reveal a regular rotation with a maximum of $\pm 90$ km/s and an elongated central velocity dispersion decrease going down to 105 km/s.
We combined the kinematics data with a luminous mass model from dust-corrected HST WFPC2 F606W and SDSS r-band images and constructed dynamical JAM and Schwarzschild models from the light model and the kinematic information. The dynamical models cannot constrain the lower mass of the central mass hole and only predict an upper limit mass. The JAM models provide $M_{\mathrm{BH}}<1.5\times 10^5 M_{\odot}$ and a M/L= $1.84 \pm 0.04 $ , while the more sophisticated orbit-based Schwarzschild models state $M_{\mathrm{BH}}< 1.56 \times 10^6 M_{\odot}$ and a M/L= $1.8 \pm 0.09$ at $3\sigma$  significance. This upper limit measurement is $1 \sigma$ below the \citet{Saglia2016} and $2.5\,\sigma$ below the \citet{Greene2016} $M_{BH}-\sigma_{\mathrm{e}}$ relation assuming $\sigma_{\mathrm{e}} = 115.5 \pm 3$ km/s. In order to analyze the robustness of the measurements, we tested how various systematic uncertainties, such as dark matter content, dust attenuation, M/L variation, distance uncertainty and kinematic tracer variation, influence the results. Taking the different uncertainties into account, we remain in the black hole uncertainty given by the $3\,\sigma$ significance of the $\chi^2$ criteria. We are able to accurately constrain the black hole upper limit to approximately five times less than the black hole mass predicted by the resolution of our instruments. This result shows that AO-supported IFU data permits us to look for black holes with masses significantly below the resolution limit. This is especially important in the low-mass black hole regime, which is still largely underpopulated in the black hole-host galaxy relations. While our measurement of NGC 4414 provides a new measurement in this undersampled black hole domain, additional black hole mass measurements are needed in order to find a consensus on what happens in the low-mass end of the scaling relations.

\begin{acknowledgements}
The authors want to thank Tony Wong for providing the CO data of NGC 4414 and Remco Van den Bosch for fruitful discussions regarding the MGE model of NGC 4414. MC acknowledges support from a Royal Society University Research Fellowship.
Based on observations obtained at the Gemini Observatory, which is operated by the Association of Universities for Research in Astronomy, Inc., under a cooperative agreement with the NSF on behalf of the Gemini partnership: the National Science Foundation (United States), the National Research Council (Canada), CONICYT (Chile), the Australian Research Council (Australia), Minist\'{e}rio da Ci\^{e}ncia, Tecnologia e Inova\c{c}\~{a}o (Brazil) and Ministerio de Ciencia, Tecnolog\'{i}a e Innovaci\'{o}n Productiva (Argentina), under program GN-2007A-Q-45. Based on observations made with the NASA/ESA Hubble Space Telescope, obtained from the ESA Hubble Science Archive at the Space Telescope Science Institute, which is operated by the Association of Universities for Research in Astronomy, Inc., under NASA contract NAS 5-26555.
Funding for SDSS-III has been provided by the Alfred P. Sloan Foundation, the Participating Institutions, the National Science Foundation, and the U.S. Department of Energy Office of Science. The SDSS-III web site is http://www.sdss3.org/.
SDSS-III is managed by the Astrophysical Research Consortium for the Participating Institutions of the SDSS-III Collaboration including the University of Arizona, the Brazilian Participation Group, Brookhaven National Laboratory, Carnegie Mellon University, University of Florida, the French Participation Group, the German Participation Group, Harvard University, the Instituto de Astrofisica de Canarias, the Michigan State/Notre Dame/JINA Participation Group, Johns Hopkins University, Lawrence Berkeley National Laboratory, Max Planck Institute for Astrophysics, Max Planck Institute for Extraterrestrial Physics, New Mexico State University, New York University, Ohio State University, Pennsylvania State University, University of Portsmouth, Princeton University, the Spanish Participation Group, University of Tokyo, University of Utah, Vanderbilt University, University of Virginia, University of Washington, and Yale University. 
This research made use of Montage. It is funded by the National Science Foundation under Grant Number ACI-1440620, and was previously funded by the National Aeronautics and Space Administration's Earth Science Technology Office, Computation Technologies Project, under Cooperative Agreement Number NCC5-626 between NASA and the California Institute of Technology.
This research has made use of the NASA/IPAC Extragalactic Database (NED) which is operated by the Jet Propulsion Laboratory, California Institute of Technology, under contract with the National Aeronautics and Space Administration. We acknowledge the usage of the HyperLeda database (http://leda.univ-lyon1.fr)
\end{acknowledgements}


\newpage
\begin{appendix}
\section{Dust correction and masking}
\subsection{SDSS/r-band image}\label{SDSSdust}
\begin{figure}[htb]
 \centering
   \includegraphics[width=\hsize]{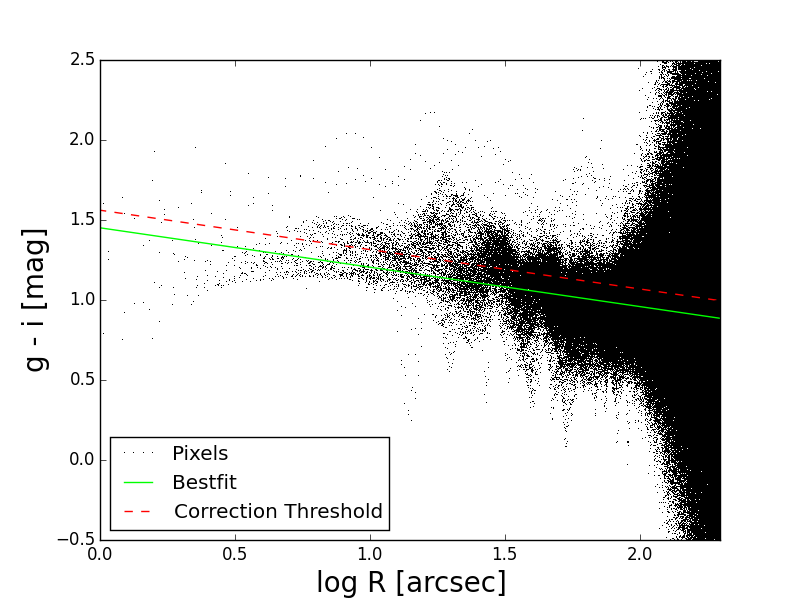}
      \caption{Color profile of the SDSS g-i map used for the dust correction. The best-fitting linear line obtained by a robust fit is shown in green. We decided the red line to be the threshold in order to assure the same width of the gradient under the line-fit and above. All pixels above the red line and for $\log(d)< 1.9\, (\approx 80 $) [arcsec] were corrected for extinction.}
         \label{dustcorrection}
\end{figure}

\begin{figure}[htb]
\centering
   \includegraphics[width=\hsize]{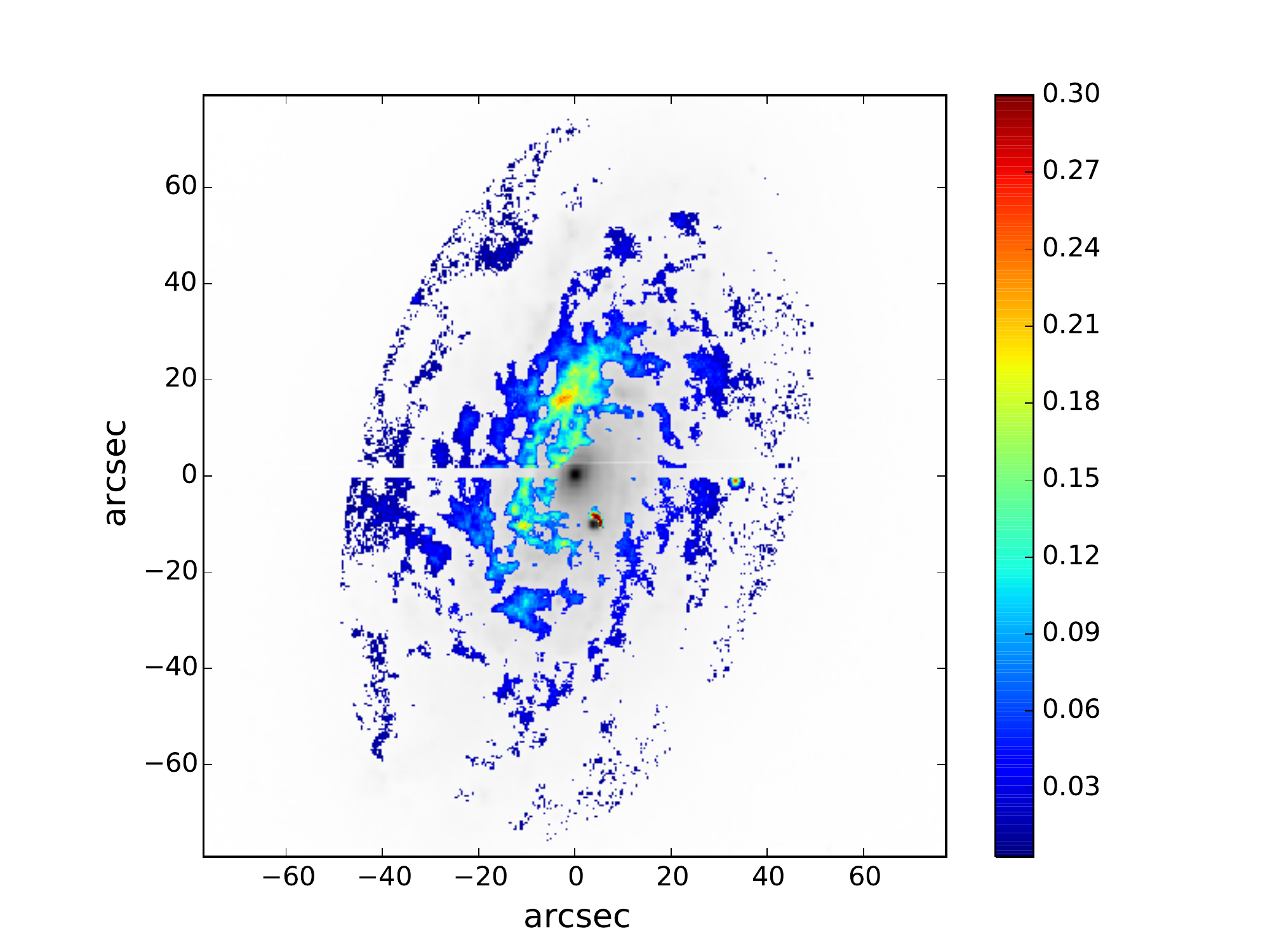}
      \caption{Dust-corrected central region of the observed SDSS r-band image. The correction was only applied within a major axis radius of $80''$ (see caption of Fig.~\ref{dustcorrection}). The over-plotted color coding indicates the degree of the correction where 0.1 means that the observed flux increased by 10 \%. The lines in the center of the image are edge artifacts from assembling the single SDSS
images to a large FOV montage.}
         \label{dustcorrection2}
\end{figure}

Before deriving the MGE model from the SDSS/r-band image, we corrected it for the effects of dust absorption based on the method described in \citet{Cappellari2002b,Scott2013}.
The main assumption of this method is that the dust can be summed up to a light absorbing screen between the observer and the galaxy \citep{Carollo1997a,Cappellari2002b}. Due to the extinction, this screen changes the intrinsic color of the dust-affected galaxy fragments which are assumed to have the same intrinsic color as the adjacent regions. According to the Galactic extinction law we derived the r-band extinction $A_r$ from the color excess between the g- and i-band images $A_r=1.15\,\mathrm{E(g-i)}$. The main steps of our dust correction were the following: 1) For each pixel, the color (g-i) was calculated and plotted over the logarithm of the semi-major axis distance; 2) Assuming that the intrinsic galaxy color varies linearly with the logarithm of the radius, we performed a robust linear fit to the radial color profile to determine the underlying color gradient of the galaxy. The color profile of NGC~\,4414 is presented in Fig.~\ref{dustcorrection}. As typically for spiral galaxies, the profile shows a color gradient with the central regions being redder than the outskirts of the galaxy. The color excess E(g-i) was then computed for each pixel as the difference between measured and intrinsic galaxy color. Thus, we retrieved a E(g-i) color excess map of NGC~\,4414. All pixels above a threshold, chosen to be E(g-i)$>0.11,$ were significantly affected by dust extinction and corrected in the SDSS/r-band image by using the Galactic extinction law. The method corrected patchy dust absorption which could be found in the disk regions of the eastern side of NGC~\,4414 which faces towards us (Fig.~\ref{dustcorrection2}). The Figure also shows quantitatively how much of the measured flux was corrected (where 0.1 means 10\%\ (blue) and 0.25 means 25\%\  (orange)). The largest correction was approximately 25\%\  of the measured flux in the eastern side of NGC 4414.
The lines in the center of the image are edge artifacts from assembling the single SDSS
images to a large FOV montage and were masked in the surface brightness modeling.

\subsection{WFPC2/F606W PC image}\label{HSTdust}
As NGC 4414 shows further dust patterns in the central regions, we also attempted to correct the dust in the F606W PC image. This dust correction is very important for the models as the PC image probes the direct vicinity of the black hole. It was not possible to apply the same dust correction as for the SDSS r-band image, as NGC 4414 PC images of other bands were all saturated in the center. Therefore, we decided for creating a dust mask in order to account for dust attenuation in the MGE modeling.
 \begin{figure}[htb]
\centering
   \includegraphics[width=0.9\hsize]{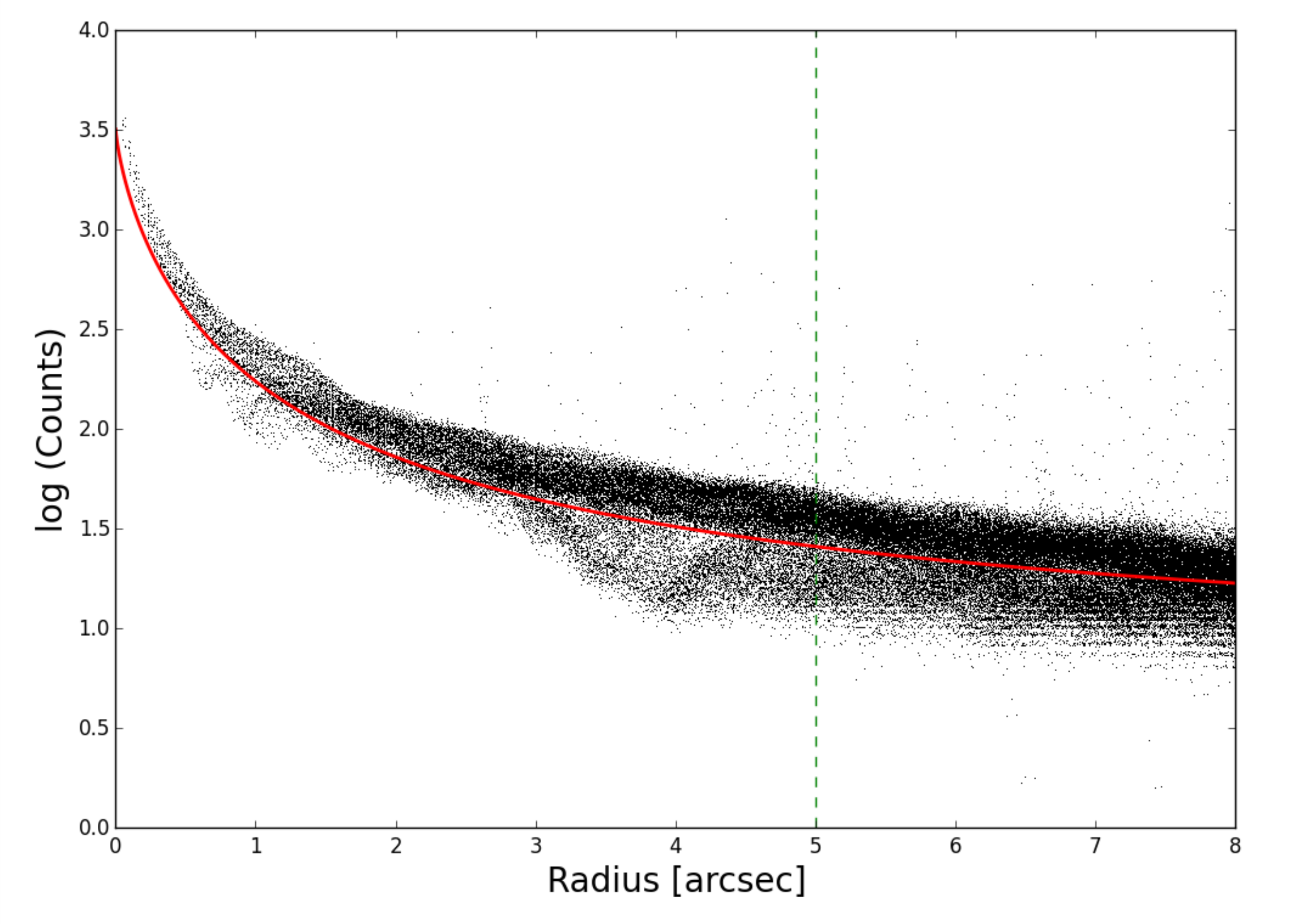}
      \caption{Logarithmic surface brightness profile (in photon counts) of the HST F606W image used for the dust-masking. Every dot is a pixel in the image, the lower envelope fit of the surface brightness is given by the red solid line. Each pixel below the line is masked in the photometric measurement of the MGE. The green dashed line marks the edge of the image which is used for the MGE modeling.}
         \label{mask_plot}
\end{figure}

\begin{figure}[htb]
\centering
   \includegraphics[width=\hsize]{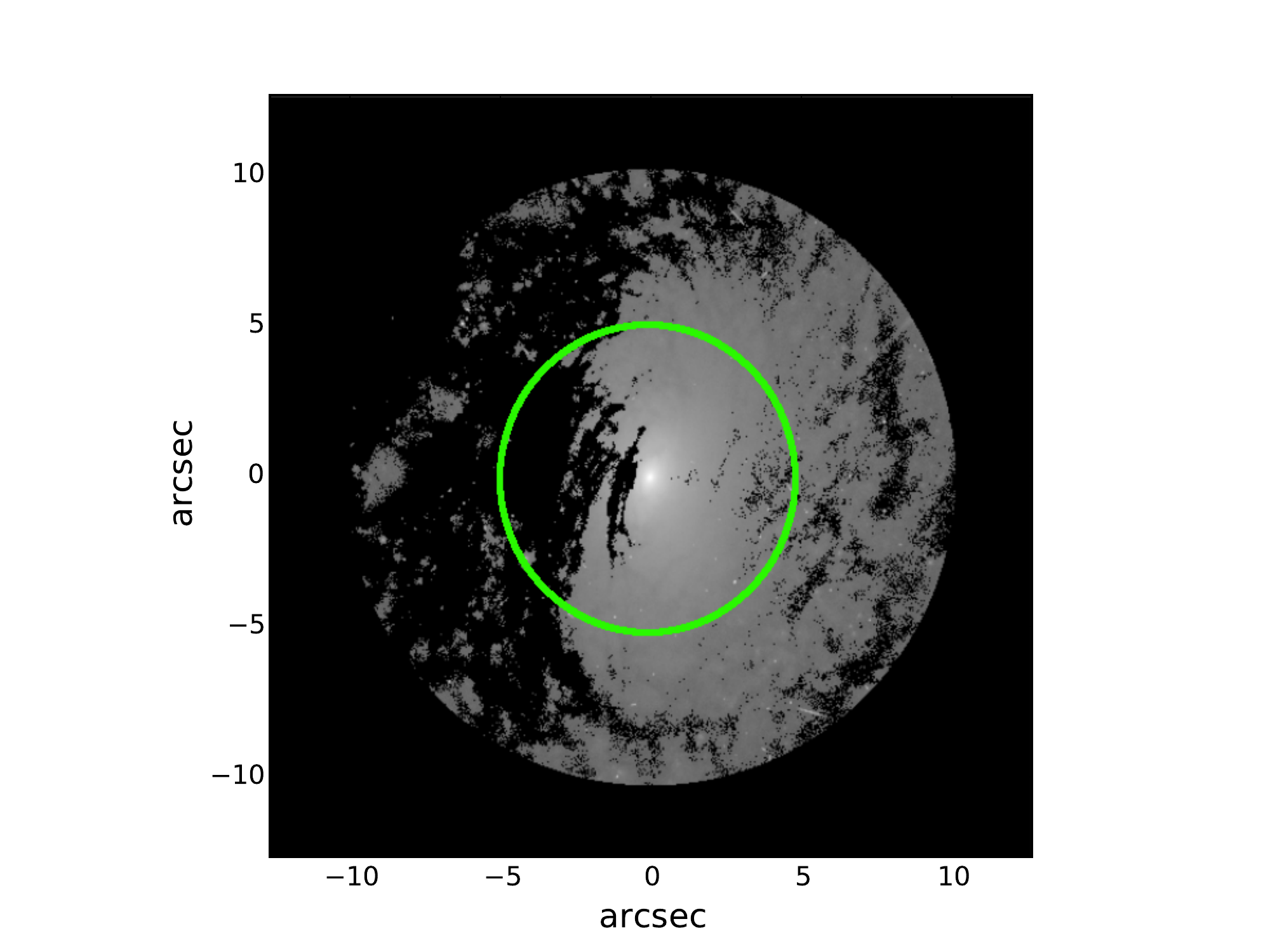}
      \caption{Dust masked region of the HST/F606W image. The black parts of the image are the regions which were masked to account for dust attenuation. The green circle marks the region which was used in the MGE modeling ($r < 5$\arcsec).}
         \label{dustcorrection3}
\end{figure}
We chose to mask those pixels which deviated significantly from the characteristic galaxy surface brightness profile (Fig.~\ref{mask_plot}). Both the dust patches and hot pixels are clearly evident in this profile. We distinguished the dust affected from the unaffected pixels by fitting the lower envelope of the main surface brightness profile with an appropriate function (four parameter logistic function) and masked the pixels below this fit. The masked pixels are shown in Fig.~\ref{dustcorrection3}. 

\section{Determination of the HST, NIFS and GMOS PSF}
\subsection{HST spatial resolution}
\label{HST PSF}
In order to compare the MGE model of NGC~\,4414 (Section~\ref{mass model}) with the observed surface brightness, it is necessary to convolve it with the central image PSF. We generated the PSF for the WFPC2/F606PC image by using Tiny Tim. The PSF was modeled by a sum of concentric circular Gaussians using the MGE method. Each of the Gaussians was assigned a relative weight which is normalized such that the sum of the  weights equals one. The MGE parameters of the single Gaussians are given in Table B.1. On the other hand, modeling the Tiny Tim PSF with one circular Gaussian, we obtained ${\rm FWHM = 0.09''}$.
\begin{table}
\caption{MGE parameters of the HST/WFPC2/F606PC circular PSF fit}
\centering
\begin{tabular}{ccc}
\hline\hline
k & $G_k$  &  $\sigma_k^*$ \\
 &   &  (arcsec) \\
\hline
1 & 0.238 & 0.0173 \\
2 & 0.569 & 0.0511 \\
3 & 0.0842 & 0.145\\
4 & 0.0683 & 0.334\\
5 & 0.0406 & 0.847\\
\hline
\end{tabular}
\end{table}
\subsection{NIFS and GMOS spatial resolution}\label{NIFS GMOS PSF}
\begin{figure}
\centering
   \includegraphics[width=0.78\hsize]{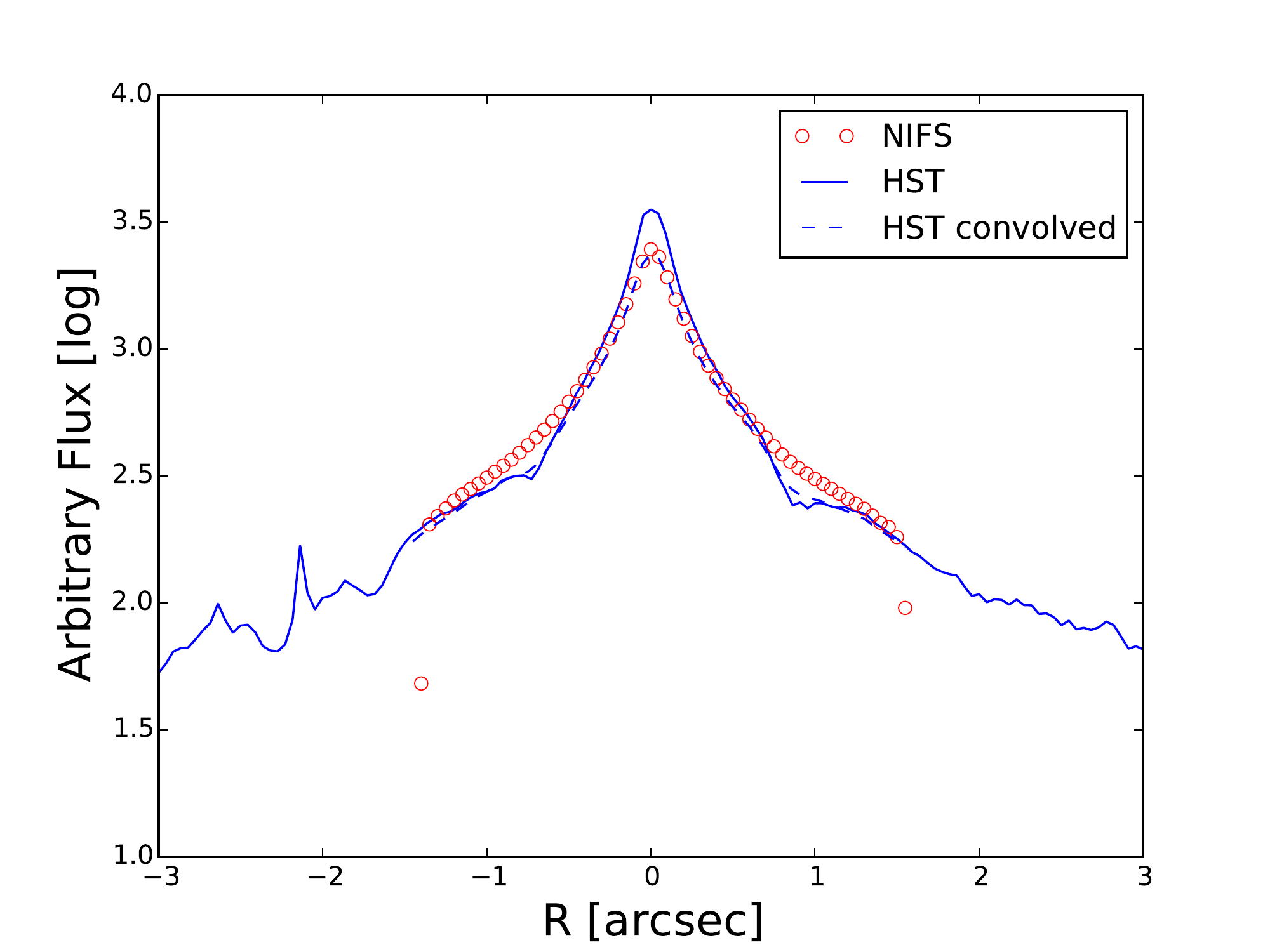}
   \includegraphics[width=0.78\hsize]{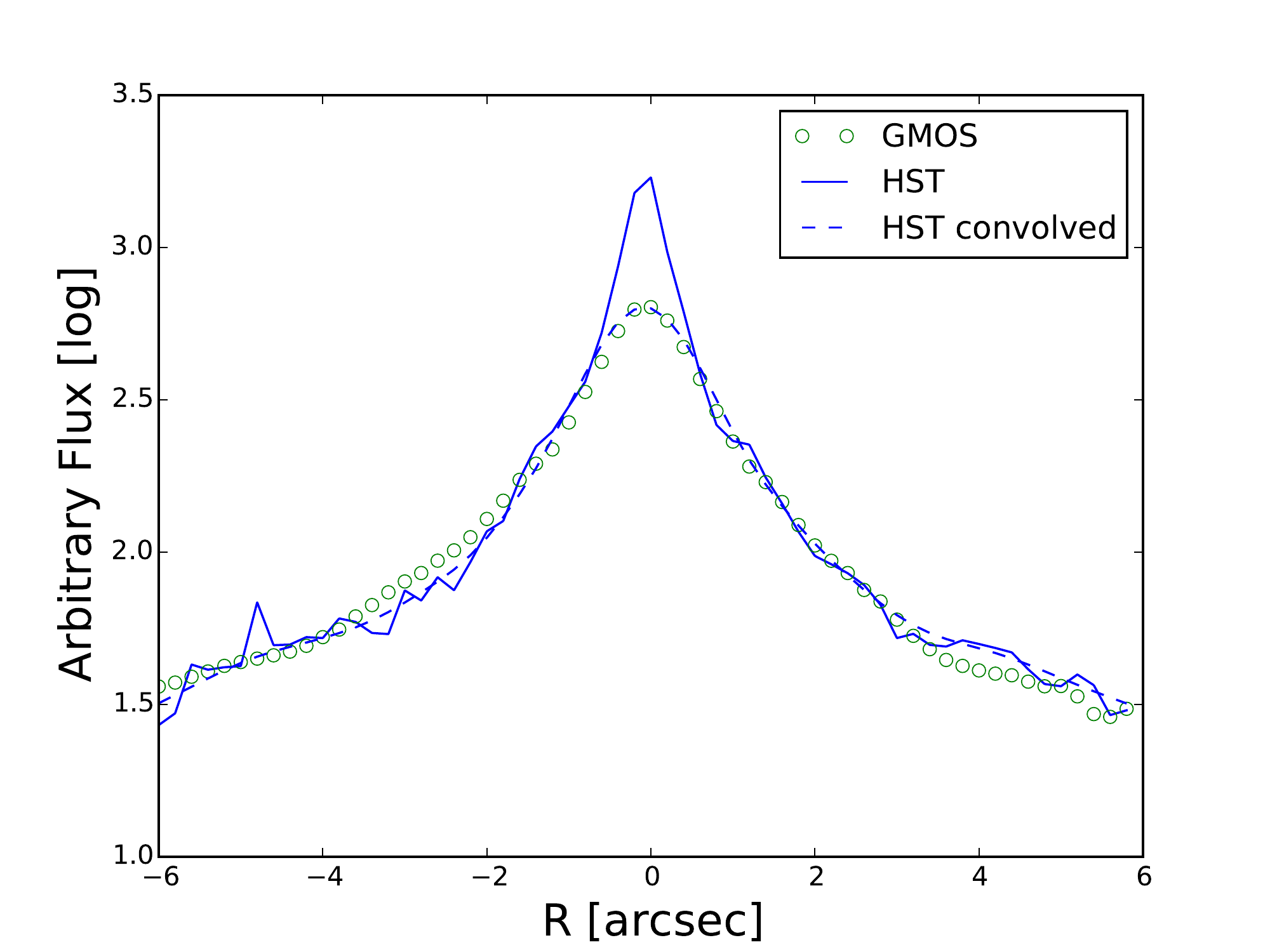}
      \caption{Determination of the NIFS (top) and GMOS (bottom) PSF by comparing the light profiles of the NIFS/GMOS and HST images along the major axis. In both panels the open circles denote the IFU profiles and blue solid lines the HST profiles. The HST profiles were convolved with the PSFs of Table~\ref{PSF parameters} to fit the IFU data and are shown as dashed blue lines. }
         \label{PSF}
\end{figure}

In order to determine to which scales we can probe the inner dynamics of NGC 4414, a proper estimate of the spatial resolution of our IFU data is indispensable. When no point sources can be found in the observed FOV, the observed images have to be compared with reference images of significantly higher resolution \citep{Davies2008,Krajnovic2009a}.
The HST/WFPC2 image that we obtained to derive our luminosity mass model provides sufficient resolution. The spatial resolution of the IFU's could then be determined by convolving the HST image with a PSF and then degrading the image until it matched the reconstructed IFU image. A good description for the PSF can be obtained by a sum of two circular and concentric Gaussians, G, one having a broad- and the other a narrow shape
\begin{equation}
{\rm G_{PSF}}= f_1 *{\rm G( FWHM_1})+ (1- f_1) *{\rm G( FWHM_2})
,\end{equation}
where $f_1$ is the relative flux of the narrow Gaussian. We used two Gaussians to determine the PSF of the NIFS observation and a single Gaussian for the GMOS observation. In order to compare the different images with each other, we aligned them by rotating them such that the major and minor axes of the galaxies were coinciding. After the convolution, the HST image was rebinned to the same pixel scale as the respective IFU observation. The best fitting PSF parameters are found by minimizing the residual between the convolved HST image and the reconstructed IFU image. Finally, the PSF of the HST image (Sect.~\ref{HST PSF}) had to be added quadratically to the measured PSFs. The best-fitting parameters of our PSFs are given in Table~\ref{PSF parameters} and a comparison of the light profiles of the used images along the major axis is shown in Fig.~\ref{PSF}. The convolved HST profiles (blue dashed lines) reproduce the IFU (open circles) profiles very well in the majority of cases.

\subsection{Strehl ratio}
\label{strehlratio}
The Strehl ratio measures the effect of wavefront aberrations on the optical quality of the observations.
It can be determined by comparing the peak intensity of the measured PSF and the peak intensity of the ideal diffraction limited PSF assuming an ideal working LGS AO. In this work, we obtained the FWHM of the narrow Gaussian component of the NIFS observation to be $0.126''$ (see table 2), while the diffraction limited FWHM at 2.2 microns on the Gemini 8 m telescope is approximately $0.07"$ \citep{McGregor1999}. We created normalized 2D Gaussians from those two FWHM and compared the peak intensities to obtain a Strehl ratio of $30.8 \%$.

\section{Comparison of Schwarzschild models and velocity moments for NIFS and GMOS}
\label{app:models}

Dynamical models do not constrain the lower mass limit for a SMBH in NGC~\,4414, with the formal best fit model being at the edge of our grid in M$_{\rm BH}$. Therefore we present here a representative model close to the upper mass limit given by the 3$\sigma$ uncertainty level. The model reproduces most of the kinematics features both on NIFS and GMOS data. For the NIFS data, the major difference is an over prediction of the central velocity dispersion and under prediction of $h_4$ Gauss-Hermite coefficient at approximately two and one sigma level. For GMOS data the models  reproduce the kinematics maps to a similarly good standard with the exception of the central regions on the velocity dispersion and the $h_4$ maps. Note that the central one arcsec was not fitted for the GMOS data, which partially explains the over-prediction of the velocity dispersion and the under-prediction of $h_4$ at approximately the two sigma level.  

 \begin{figure*}[htb]
\centering
   \includegraphics[width=0.7\hsize]{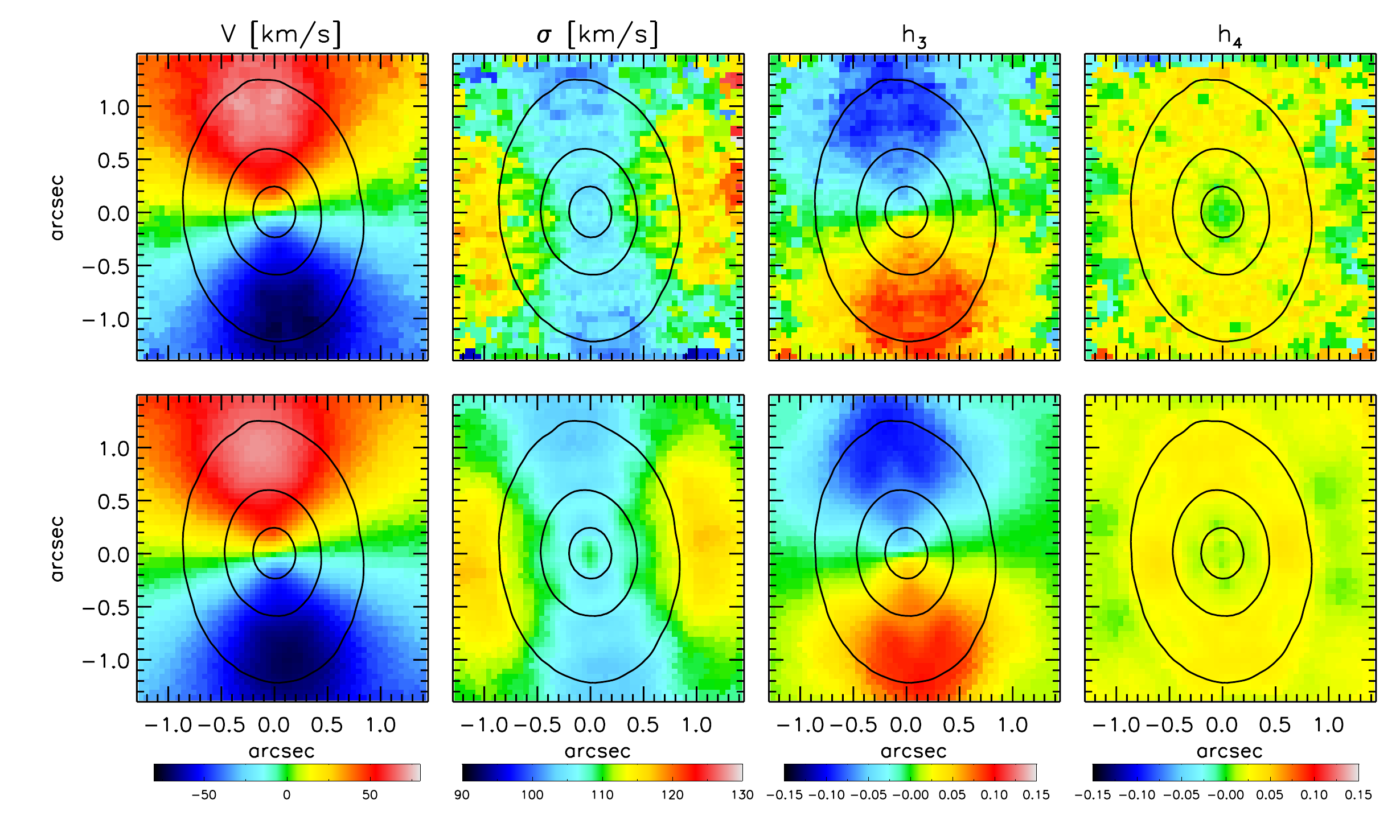}
   \includegraphics[width=0.7\hsize]{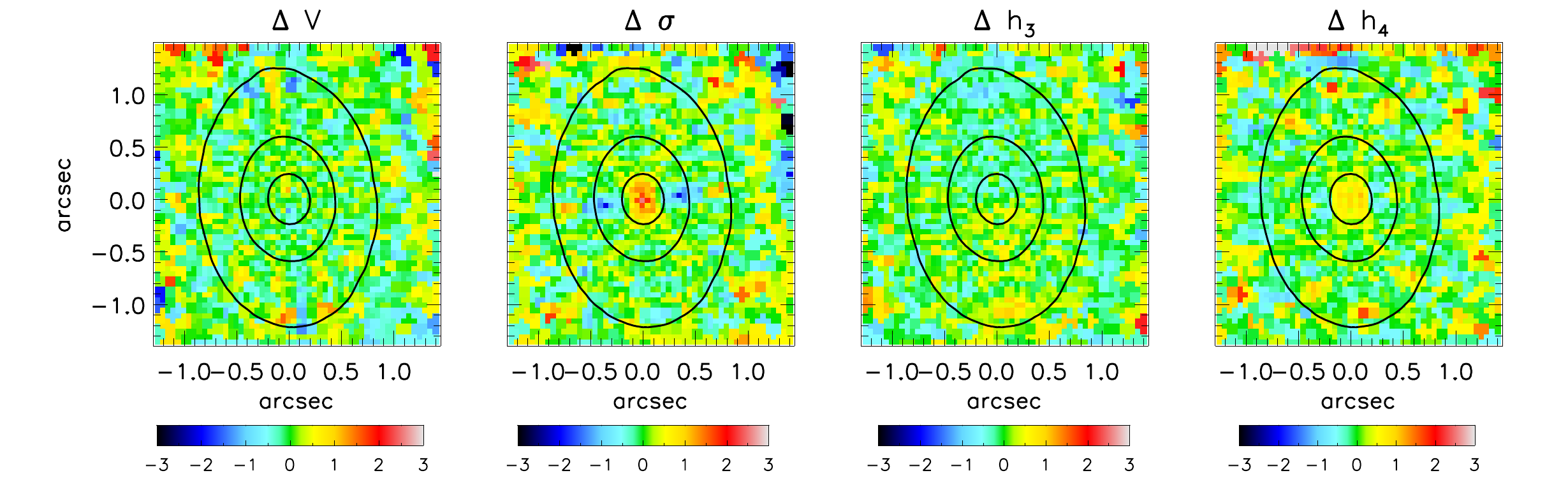}
      \caption{Top row: Symmetrized NIFS maps showing (from left to right) the mean velocity, the velocity dispersion and $h_3$ and $h_4$ Gauss-Hermite moments. Middle row: Schwarzschild dynamical models for M/L$ =1.80$ and M$_{\rm BH} = 5.81\times10^5$ M$_\odot$. The maps show the same quantities as in the row above. Bottom row: the difference between the Schwarzschild model predicted and the observed kinematics, divided by the observational errors. }
         \label{App:NIFS}
\end{figure*}

 \begin{figure*}[htb]
\centering
   \includegraphics[width=0.8\hsize]{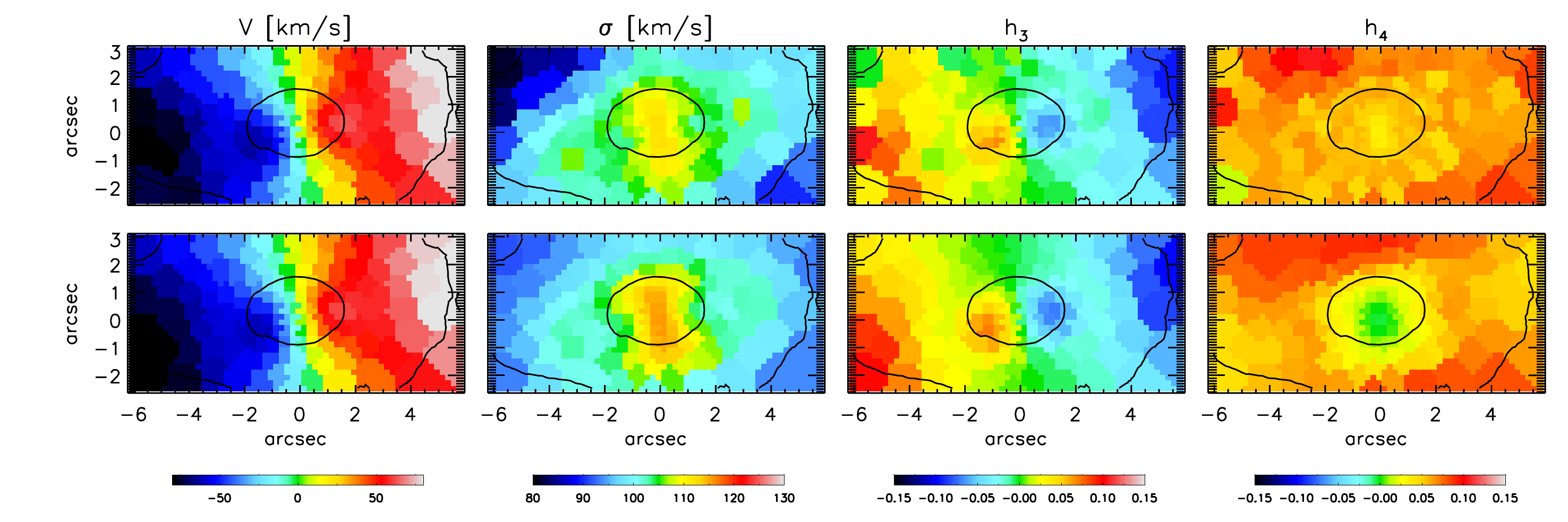}
   \includegraphics[width=0.8\hsize]{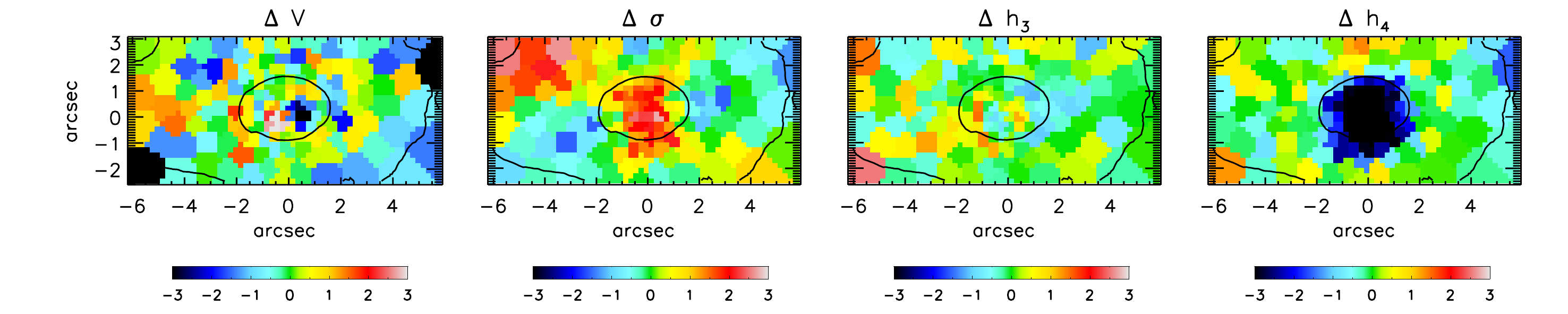}
      \caption{As in Fig.~\ref{App:NIFS} but for GMOS data.}
         \label{App:GMOS}
\end{figure*}

\end{appendix}

\begin{thebibliography}{108}
\expandafter\ifx\csname natexlab\endcsname\relax\def\natexlab#1{#1}\fi

\bibitem[{{Allington-Smith} {et~al.}(2002){Allington-Smith}, {Murray},
  {Content}, {Dodsworth}, {Davies}, {Miller}, {Turner}, {Jorgensen}, {Hook},
  {Crampton}, \& {Murowinski}}]{Allington-Smith2002}
{Allington-Smith}, J., {Murray}, G., {Content}, R., {et~al.} 2002, in
  Astronomical Society of the Pacific Conference Series, Vol. 282, Galaxies:
  the Third Dimension, ed. M.~{Rosada}, L.~{Binette}, \& L.~{Arias}, 415

\bibitem[{{Atkinson} {et~al.}(2005){Atkinson}, {Collett}, {Marconi}, {Axon},
  {Alonso-Herrero}, {Batcheldor}, {Binney}, {Capetti}, {Carollo}, {Dressel},
  {Ford}, {Gerssen}, {Hughes}, {Macchetto}, {Maciejewski}, {Merrifield},
  {Scarlata}, {Sparks}, {Stiavelli}, {Tsvetanov}, \& {van der
  Marel}}]{Atkinson2005}
{Atkinson}, J.~W., {Collett}, J.~L., {Marconi}, A., {et~al.} 2005, \mnras, 359,
  504

\bibitem[{Baldassare {et~al.}(2015)Baldassare, Reines, Gallo, \&
  Greene}]{Baldassare2015}
Baldassare, V.~F., Reines, A.~E., Gallo, E., \& Greene, J.~E. 2015, The
  Astrophysical Journal, 809, L14

\bibitem[{Barth {et~al.}(2002)Barth, Ho, \& Sargent}]{Barth2002}
Barth, A.~J., Ho, L.~C., \& Sargent, W. L.~W. 2002, The Astronomical Journal,
  124, 2607–2614

\bibitem[{{Beifiori} {et~al.}(2011){Beifiori}, {Maraston}, {Thomas}, \&
  {Johansson}}]{Beifiori2011}
{Beifiori}, A., {Maraston}, C., {Thomas}, D., \& {Johansson}, J. 2011, \aap,
  531, A109

\bibitem[{{Bell} \& {de Jong}(2001)}]{Bell2001}
{Bell}, E.~F. \& {de Jong}, R.~S. 2001, \apj, 550, 212

\bibitem[{{Bender} {et~al.}(1994){Bender}, {Saglia}, \& {Gerhard}}]{Bender1994}
{Bender}, R., {Saglia}, R.~P., \& {Gerhard}, O.~E. 1994, \mnras, 269, 785

\bibitem[{{Bentz} {et~al.}(2016){Bentz}, {Batiste}, {Seals}, {Garcia}, {Kuzio
  de Naray}, {Peters}, {Anderson}, {Jones}, {Lester}, {Machuca}, {Parks},
  {Pope}, {Revalski}, {Roberts}, {Saylor}, {Sevrinsky}, \&
  {Turner}}]{Bentz2016}
{Bentz}, M.~C., {Batiste}, M., {Seals}, J., {et~al.} 2016, ArXiv e-prints

\bibitem[{{Binney} \& {Tremaine}(2008)}]{Binney2008}
{Binney}, J. \& {Tremaine}, S. 2008, {Galactic Dynamics: Second Edition}
  (Princeton University Press)

\bibitem[{{Braine} {et~al.}(1997){Braine}, {Brouillet}, \&
  {Baudry}}]{Braine1997}
{Braine}, J., {Brouillet}, N., \& {Baudry}, A. 1997, \aap, 318, 19

\bibitem[{Cappellari(2002)}]{Cappellari2002}
Cappellari, M. 2002, \mnras, 333, 400–410

\bibitem[{{Cappellari}(2008)}]{Cappellari2008}
{Cappellari}, M. 2008, \mnras, 390, 71

\bibitem[{{Cappellari} {et~al.}(2006){Cappellari}, {Bacon}, {Bureau}, {Damen},
  {Davies}, {de Zeeuw}, {Emsellem}, {Falc{\'o}n-Barroso}, {Krajnovi{\'c}},
  {Kuntschner}, {McDermid}, {Peletier}, {Sarzi}, {van den Bosch}, \& {van de
  Ven}}]{Cappellari2006}
{Cappellari}, M., {Bacon}, R., {Bureau}, M., {et~al.} 2006, \mnras, 366, 1126

\bibitem[{{Cappellari} \& {Copin}(2003)}]{Cappellari2003}
{Cappellari}, M. \& {Copin}, Y. 2003, \mnras, 342, 345

\bibitem[{{Cappellari} \& {Emsellem}(2004)}]{Cappellari2004}
{Cappellari}, M. \& {Emsellem}, E. 2004, \pasp, 116, 138

\bibitem[{{Cappellari} {et~al.}(2007){Cappellari}, {Emsellem}, {Bacon},
  {Bureau}, {Davies}, {de Zeeuw}, {Falc{\'o}n-Barroso}, {Krajnovi{\'c}},
  {Kuntschner}, {McDermid}, {Peletier}, {Sarzi}, {van den Bosch}, \& {van de
  Ven}}]{Cappellari2007}
{Cappellari}, M., {Emsellem}, E., {Bacon}, R., {et~al.} 2007, \mnras, 379, 418

\bibitem[{{Cappellari} {et~al.}(2013{\natexlab{a}}){Cappellari}, {McDermid},
  {Alatalo}, {Blitz}, {Bois}, {Bournaud}, {Bureau}, {Crocker}, {Davies},
  {Davis}, {de Zeeuw}, {Duc}, {Emsellem}, {Khochfar}, {Krajnovi{\'c}},
  {Kuntschner}, {Morganti}, {Naab}, {Oosterloo}, {Sarzi}, {Scott}, {Serra},
  {Weijmans}, \& {Young}}]{Cappellari2013b}
{Cappellari}, M., {McDermid}, R.~M., {Alatalo}, K., {et~al.}
  2013{\natexlab{a}}, \mnras, 432, 1862

\bibitem[{{Cappellari} {et~al.}(2010){Cappellari}, {McDermid}, {Bacon},
  {Davies}, {de Zeeuw}, {Emsellem}, {Falc{\'o}n-Barroso}, {Krajnovi{\'c}},
  {Kuntschner}, {Peletier}, {Sarzi}, {van den Bosch}, \& {van de
  Ven}}]{Cappellari2010}
{Cappellari}, M., {McDermid}, R.~M., {Bacon}, R., {et~al.} 2010, in American
  Institute of Physics Conference Series, Vol. 1240, American Institute of
  Physics Conference Series, ed. V.~P. {Debattista} \& C.~C. {Popescu},
  211--214

\bibitem[{{Cappellari} {et~al.}(2013{\natexlab{b}}){Cappellari}, {Scott},
  {Alatalo}, {Blitz}, {Bois}, {Bournaud}, {Bureau}, {Crocker}, {Davies},
  {Davis}, {de Zeeuw}, {Duc}, {Emsellem}, {Khochfar}, {Krajnovi{\'c}},
  {Kuntschner}, {McDermid}, {Morganti}, {Naab}, {Oosterloo}, {Sarzi}, {Serra},
  {Weijmans}, \& {Young}}]{Cappellari2013}
{Cappellari}, M., {Scott}, N., {Alatalo}, K., {et~al.} 2013{\natexlab{b}},
  \mnras, 432, 1709

\bibitem[{{Cappellari} {et~al.}(2002){Cappellari}, {Verolme}, {van der Marel},
  {Verdoes Kleijn}, {Illingworth}, {Franx}, {Carollo}, \& {de
  Zeeuw}}]{Cappellari2002b}
{Cappellari}, M., {Verolme}, E.~K., {van der Marel}, R.~P., {et~al.} 2002,
  \apj, 578, 787

\bibitem[{{Carollo} {et~al.}(1997){Carollo}, {Franx}, {Illingworth}, \&
  {Forbes}}]{Carollo1997a}
{Carollo}, C.~M., {Franx}, M., {Illingworth}, G.~D., \& {Forbes}, D.~A. 1997,
  \apj, 481, 710

\bibitem[{Cleveland(1979)}]{Cleveland1979}
Cleveland, W.~S. 1979, Journal of the American Statistical Association, 74, 829

\bibitem[{Cleveland \& Devlin(1988)}]{Cleveland1988}
Cleveland, W.~S. \& Devlin, S.~J. 1988, Journal of the American Statistical
  Association, 83, 596

\bibitem[{{Coccato} {et~al.}(2006){Coccato}, {Sarzi}, {Pizzella}, {Corsini},
  {Dalla Bont{\`a}}, \& {Bertola}}]{Coccato2006}
{Coccato}, L., {Sarzi}, M., {Pizzella}, A., {et~al.} 2006, \mnras, 366, 1050

\bibitem[{Cretton \& van~den Bosch(1999)}]{Cretton1999}
Cretton, N. \& van~den Bosch, F.~C. 1999, The Astrophysical Journal, 514,
  704–724

\bibitem[{{Davies}(2008)}]{Davies2008}
{Davies}, R. 2008, in 2007 ESO Instrument Calibration Workshop, ed. A.~{Kaufer}
  \& F.~{Kerber}, 249

\bibitem[{Davis {et~al.}(2012)Davis, Alatalo, Bureau, Cappellari, Scott, Young,
  Blitz, Crocker, Bayet, Bois, \& et~al.}]{Davis2012}
Davis, T.~A., Alatalo, K., Bureau, M., {et~al.} 2012, Monthly Notices of the
  Royal Astronomical Society, 429, 534–555

\bibitem[{{de Blok} {et~al.}(2014){de Blok}, {J{\'o}zsa}, {Patterson},
  {Gentile}, {Heald}, {J{\"u}tte}, {Kamphuis}, {Rand}, {Serra}, \&
  {Walterbos}}]{deBlok2014}
{de Blok}, W.~J.~G., {J{\'o}zsa}, G.~I.~G., {Patterson}, M., {et~al.} 2014,
  \aap, 566, A80

\bibitem[{{De Lorenzi} {et~al.}(2013){De Lorenzi}, {Hartmann}, {Debattista},
  {Seth}, \& {Gerhard}}]{DeLorenzi2013}
{De Lorenzi}, F., {Hartmann}, M., {Debattista}, V.~P., {Seth}, A.~C., \&
  {Gerhard}, O. 2013, \mnras, 429, 2974

\bibitem[{{de Vaucouleurs} {et~al.}(1991){de Vaucouleurs}, {de Vaucouleurs},
  {Corwin}, {Buta}, {Paturel}, \& {Fouqu{\'e}}}]{deVaucouleurs1991rc3}
{de Vaucouleurs}, G., {de Vaucouleurs}, A., {Corwin}, Jr., H.~G., {et~al.}
  1991, {Third Reference Catalogue of Bright Galaxies. Volume I: Explanations
  and references. Volume II: Data for galaxies between 0$^{h}$ and 12$^{h}$.
  Volume III: Data for galaxies between 12$^{h}$ and 24$^{h}$.}

\bibitem[{{den Brok} {et~al.}(2015){den Brok}, {Seth}, {Barth}, {Carson},
  {Neumayer}, {Cappellari}, {Debattista}, {Ho}, {Hood}, \&
  {McDermid}}]{denBrok2015}
{den Brok}, M., {Seth}, A.~C., {Barth}, A.~J., {et~al.} 2015, \apj, 809, 101

\bibitem[{{Drehmer} {et~al.}(2015){Drehmer}, {Storchi-Bergmann}, {Ferrari},
  {Cappellari}, \& {Riffel}}]{Drehmer2015}
{Drehmer}, D.~A., {Storchi-Bergmann}, T., {Ferrari}, F., {Cappellari}, M., \&
  {Riffel}, R.~A. 2015, \mnras, 450, 128

\bibitem[{{Emsellem}(2013)}]{Emsellem2013}
{Emsellem}, E. 2013, \mnras, 433, 1862

\bibitem[{{Emsellem} {et~al.}(1994){Emsellem}, {Monnet}, \&
  {Bacon}}]{Emsellem1994}
{Emsellem}, E., {Monnet}, G., \& {Bacon}, R. 1994, \aap, 285

\bibitem[{{Falc{\'o}n-Barroso} {et~al.}(2016){Falc{\'o}n-Barroso}, {Lyubenova},
  {van de Ven}, {M{\'e}ndez-Abreu}, {Aguerri}, {Garc{\'{\i}}a-Lorenzo},
  {Bekeraite}, {S{\'a}nchez}, {Husemann}, {Garc{\'{\i}}a-Benito}, {Mast},
  {Walcher}, {Zibetti}, {Barrera-Ballesteros}, {Galbany},
  {S{\'a}nchez-Bl{\'a}zquez}, {Singh}, {van den Bosch}, {Wild}, {Zhu},
  {Bland-Hawthorn}, {Cid Fernandes}, {de Lorenzo-C{\'a}ceres}, {Gallazzi},
  {Gonz{\'a}lez Delgado}, {Marino}, {M{\'a}rquez}, {P{\'e}rez}, {P{\'e}rez},
  {Roth}, {Rosales-Ortega}, {Ru{\'{\i}}z-Lara}, {Wisotzki}, {Ziegler}, \& {the
  CALIFA collaboration}}]{Falcon-Barroso2016b}
{Falc{\'o}n-Barroso}, J., {Lyubenova}, M., {van de Ven}, G., {et~al.} 2016,
  ArXiv e-prints

\bibitem[{{Falc{\'o}n-Barroso} {et~al.}(2011){Falc{\'o}n-Barroso},
  {S{\'a}nchez-Bl{\'a}zquez}, {Vazdekis}, {Ricciardelli}, {Cardiel}, {Cenarro},
  {Gorgas}, \& {Peletier}}]{Falcon-Barroso2011}
{Falc{\'o}n-Barroso}, J., {S{\'a}nchez-Bl{\'a}zquez}, P., {Vazdekis}, A.,
  {et~al.} 2011, \aap, 532, A95

\bibitem[{{Ferrarese} \& {Ford}(2005)}]{Ferrarese2005}
{Ferrarese}, L. \& {Ford}, H. 2005, \ssr, 116, 523

\bibitem[{{Ferrarese} \& {Merritt}(2000)}]{Ferrarese2000}
{Ferrarese}, L. \& {Merritt}, D. 2000, \apjl, 539, L9

\bibitem[{{Fisher} {et~al.}(2009){Fisher}, {Drory}, \&
  {Fabricius}}]{Fisher2009}
{Fisher}, D.~B., {Drory}, N., \& {Fabricius}, M.~H. 2009, \apj, 697, 630

\bibitem[{{Freedman} {et~al.}(2001){Freedman}, {Madore}, {Gibson}, {Ferrarese},
  {Kelson}, {Sakai}, {Mould}, {Kennicutt}, {Ford}, {Graham}, {Huchra},
  {Hughes}, {Illingworth}, {Macri}, \& {Stetson}}]{Freedman2001}
{Freedman}, W.~L., {Madore}, B.~F., {Gibson}, B.~K., {et~al.} 2001, \apj, 553,
  47

\bibitem[{{Gebhardt} {et~al.}(2000){Gebhardt}, {Bender}, {Bower}, {Dressler},
  {Faber}, {Filippenko}, {Green}, {Grillmair}, {Ho}, {Kormendy}, {Lauer},
  {Magorrian}, {Pinkney}, {Richstone}, \& {Tremaine}}]{Gebhardt2000}
{Gebhardt}, K., {Bender}, R., {Bower}, G., {et~al.} 2000, \apjl, 539, L13

\bibitem[{{Gebhardt} {et~al.}(2001){Gebhardt}, {Lauer}, {Kormendy}, {Pinkney},
  {Bower}, {Green}, {Gull}, {Hutchings}, {Kaiser}, {Nelson}, {Richstone}, \&
  {Weistrop}}]{Gebhardt2001}
{Gebhardt}, K., {Lauer}, T.~R., {Kormendy}, J., {et~al.} 2001, \aj, 122, 2469

\bibitem[{{Gebhardt} {et~al.}(2003){Gebhardt}, {Richstone}, {Tremaine},
  {Lauer}, {Bender}, {Bower}, {Dressler}, {Faber}, {Filippenko}, {Green},
  {Grillmair}, {Ho}, {Kormendy}, {Magorrian}, \& {Pinkney}}]{Gebhardt2003}
{Gebhardt}, K., {Richstone}, D., {Tremaine}, S., {et~al.} 2003, \apj, 583, 92

\bibitem[{{Gebhardt} \& {Thomas}(2009)}]{Gebhardt2009}
{Gebhardt}, K. \& {Thomas}, J. 2009, \apj, 700, 1690

\bibitem[{{Gerhard}(1993)}]{Gerhard1993}
{Gerhard}, O.~E. 1993, \mnras, 265, 213

\bibitem[{{Graham}(2016)}]{Graham2016}
{Graham}, A.~W. 2016, Galactic Bulges, 418, 263

\bibitem[{{Graham} {et~al.}(2011){Graham}, {Onken}, {Athanassoula}, \&
  {Combes}}]{Graham2011}
{Graham}, A.~W., {Onken}, C.~A., {Athanassoula}, E., \& {Combes}, F. 2011,
  \mnras, 412, 2211

\bibitem[{{Greene} {et~al.}(2010){Greene}, {Peng}, {Kim}, {Kuo}, {Braatz},
  {Impellizzeri}, {Condon}, {Lo}, {Henkel}, \& {Reid}}]{Greene2010}
{Greene}, J.~E., {Peng}, C.~Y., {Kim}, M., {et~al.} 2010, \apj, 721, 26

\bibitem[{{Greene} {et~al.}(2016){Greene}, {Seth}, {Kim}, {Laesker},
  {Goulding}, {Gao}, {Braatz}, {Henkel}, {Condon}, {Lo}, \&
  {Zhao}}]{Greene2016}
{Greene}, J.~E., {Seth}, A.~C., {Kim}, M., {et~al.} 2016, ArXiv e-prints

\bibitem[{{G{\"u}ltekin} {et~al.}(2009){G{\"u}ltekin}, {Richstone}, {Gebhardt},
  {Lauer}, {Tremaine}, {Aller}, {Bender}, {Dressler}, {Faber}, {Filippenko},
  {Green}, {Ho}, {Kormendy}, {Magorrian}, {Pinkney}, \&
  {Siopis}}]{Gueltekin2009}
{G{\"u}ltekin}, K., {Richstone}, D.~O., {Gebhardt}, K., {et~al.} 2009, \apj,
  698, 198

\bibitem[{Ho {et~al.}(2009)Ho, Greene, Filippenko, \& Sargent}]{Ho2009}
Ho, L.~C., Greene, J.~E., Filippenko, A.~V., \& Sargent, W. L.~W. 2009, The
  Astrophysical Journal Supplement Series, 183, 1–16

\bibitem[{{Jeans}(1922)}]{Jeans1922}
{Jeans}, J.~H. 1922, \mnras, 82, 122

\bibitem[{{Kanbur} {et~al.}(2003){Kanbur}, {Ngeow}, {Nikolaev}, {Tanvir}, \&
  {Hendry}}]{Kanbur2003}
{Kanbur}, S.~M., {Ngeow}, C., {Nikolaev}, S., {Tanvir}, N.~R., \& {Hendry},
  M.~A. 2003, \aap, 411, 361

\bibitem[{Kormendy \& Ho(2013)}]{kormendy2013}
Kormendy, J. \& Ho, L.~C. 2013, Annu. Rev. Astro. Astrophys., 51, 511–653

\bibitem[{{Krajnovi{\'c}} {et~al.}(2005){Krajnovi{\'c}}, {Cappellari},
  {Emsellem}, {McDermid}, \& {de Zeeuw}}]{Krajnovic2005}
{Krajnovi{\'c}}, D., {Cappellari}, M., {Emsellem}, E., {McDermid}, R.~M., \&
  {de Zeeuw}, P.~T. 2005, \mnras, 357, 1113

\bibitem[{{Krajnovi{\'c}} {et~al.}(2009){Krajnovi{\'c}}, {McDermid},
  {Cappellari}, \& {Davies}}]{Krajnovic2009a}
{Krajnovi{\'c}}, D., {McDermid}, R.~M., {Cappellari}, M., \& {Davies}, R.~L.
  2009, \mnras, 399, 1839

\bibitem[{{Krist} \& {Hook}(2001)}]{Krist2001}
{Krist}, J. \& {Hook}, R. 2001, {The Tiny Tim User's Manual, version 6.3}

\bibitem[{{Kroupa}(2001)}]{Kroupa2001}
{Kroupa}, P. 2001, \mnras, 322, 231

\bibitem[{{Lawson} \& {Hanson}(1974)}]{Lawson1974}
{Lawson}, C.~L. \& {Hanson}, R.~J. 1974, {Solving least squares problems}

\bibitem[{{McConnell} {et~al.}(2013){McConnell}, {Chen}, {Ma}, {Greene},
  {Lauer}, \& {Gebhardt}}]{McConnell2013b}
{McConnell}, N.~J., {Chen}, S.-F.~S., {Ma}, C.-P., {et~al.} 2013, \apjl, 768,
  L21

\bibitem[{{McConnell} \& {Ma}(2013)}]{McConnell2013}
{McConnell}, N.~J. \& {Ma}, C.-P. 2013, \apj, 764, 184

\bibitem[{{McDermid} {et~al.}(2015){McDermid}, {Alatalo}, {Blitz}, {Bournaud},
  {Bureau}, {Cappellari}, {Crocker}, {Davies}, {Davis}, {de Zeeuw}, {Duc},
  {Emsellem}, {Khochfar}, {Krajnovi{\'c}}, {Kuntschner}, {Morganti}, {Naab},
  {Oosterloo}, {Sarzi}, {Scott}, {Serra}, {Weijmans}, \&
  {Young}}]{McDermid2015}
{McDermid}, R.~M., {Alatalo}, K., {Blitz}, L., {et~al.} 2015, \mnras, 448, 3484

\bibitem[{{McDermid} {et~al.}(2006){McDermid}, {Emsellem}, {Shapiro}, {Bacon},
  {Bureau}, {Cappellari}, {Davies}, {de Zeeuw}, {Falc{\'o}n-Barroso},
  {Krajnovi{\'c}}, {Kuntschner}, {Peletier}, \& {Sarzi}}]{McDermid2006}
{McDermid}, R.~M., {Emsellem}, E., {Shapiro}, K.~L., {et~al.} 2006, \mnras,
  373, 906

\bibitem[{McGregor {et~al.}(1999)McGregor, Conroy, Bloxham, \& van
  Harmelen}]{McGregor1999}
McGregor, P.~J., Conroy, P., Bloxham, G., \& van Harmelen, J. 1999, Publ.
  Astron. Soc. Aust, 16, 273–287

\bibitem[{{McGregor} {et~al.}(2003){McGregor}, {Hart}, {Conroy}, {Pfitzner},
  {Bloxham}, {Jones}, {Downing}, {Dawson}, {Young}, {Jarnyk}, \& {Van
  Harmelen}}]{McGregor2003}
{McGregor}, P.~J., {Hart}, J., {Conroy}, P.~G., {et~al.} 2003, in \procspie,
  Vol. 4841, Instrument Design and Performance for Optical/Infrared
  Ground-based Telescopes, ed. M.~{Iye} \& A.~F.~M. {Moorwood}, 1581--1591

\bibitem[{{Merritt} {et~al.}(2001){Merritt}, {Ferrarese}, \&
  {Joseph}}]{Merritt2001}
{Merritt}, D., {Ferrarese}, L., \& {Joseph}, C.~L. 2001, Science, 293, 1116

\bibitem[{{Monnet} {et~al.}(1992){Monnet}, {Bacon}, \& {Emsellem}}]{Monnet1992}
{Monnet}, G., {Bacon}, R., \& {Emsellem}, E. 1992, \aap, 253, 366

\bibitem[{{Navarro} {et~al.}(1996){Navarro}, {Frenk}, \& {White}}]{Navarro1996}
{Navarro}, J.~F., {Frenk}, C.~S., \& {White}, S.~D.~M. 1996, \apj, 462, 563

\bibitem[{{Onken} {et~al.}(2014){Onken}, {Valluri}, {Brown}, {McGregor},
  {Peterson}, {Bentz}, {Ferrarese}, {Pogge}, {Vestergaard}, {Storchi-Bergmann},
  \& {Riffel}}]{Onken2014}
{Onken}, C.~A., {Valluri}, M., {Brown}, J.~S., {et~al.} 2014, \apj, 791, 37

\bibitem[{{Pastorini} {et~al.}(2007){Pastorini}, {Marconi}, {Capetti}, {Axon},
  {Alonso-Herrero}, {Atkinson}, {Batcheldor}, {Carollo}, {Collett}, {Dressel},
  {Hughes}, {Macchetto}, {Maciejewski}, {Sparks}, \& {van der
  Marel}}]{Pastorini2007}
{Pastorini}, G., {Marconi}, A., {Capetti}, A., {et~al.} 2007, \aap, 469, 405

\bibitem[{{Paturel} {et~al.}(2002){Paturel}, {Theureau}, {Fouqu{\'e}}, {Terry},
  {Musella}, \& {Ekholm}}]{Paturel2002}
{Paturel}, G., {Theureau}, G., {Fouqu{\'e}}, P., {et~al.} 2002, \aap, 383, 398

\bibitem[{{Portinari} \& {Salucci}(2010)}]{Portinari2010}
{Portinari}, L. \& {Salucci}, P. 2010, \aap, 521, A82

\bibitem[{{Press}(2007)}]{Press2007}
{Press}, W.~H. 2007, Numerical recipes : the art of scientific computing, 3rd
  edn. (Cambridge University Press)

\bibitem[{{Qian} {et~al.}(1995){Qian}, {de Zeeuw}, {van der Marel}, \&
  {Hunter}}]{Qian1995}
{Qian}, E.~E., {de Zeeuw}, P.~T., {van der Marel}, R.~P., \& {Hunter}, C. 1995,
  \mnras, 274, 602

\bibitem[{Reines {et~al.}(2013)Reines, Greene, \& Geha}]{Reines2013}
Reines, A.~E., Greene, J.~E., \& Geha, M. 2013, The Astrophysical Journal, 775,
  116

\bibitem[{Reines {et~al.}(2011)Reines, Sivakoff, Johnson, \&
  Brogan}]{Reines2011}
Reines, A.~E., Sivakoff, G.~R., Johnson, K.~E., \& Brogan, C.~L. 2011, Nature,
  470, 66–68

\bibitem[{{Richstone} \& {Tremaine}(1988)}]{Richstone1988}
{Richstone}, D.~O. \& {Tremaine}, S. 1988, \apj, 327, 82

\bibitem[{{Rix} {et~al.}(1997){Rix}, {de Zeeuw}, {Cretton}, {van der Marel}, \&
  {Carollo}}]{Rix1997}
{Rix}, H.-W., {de Zeeuw}, P.~T., {Cretton}, N., {van der Marel}, R.~P., \&
  {Carollo}, C.~M. 1997, \apj, 488, 702

\bibitem[{{Rusli} {et~al.}(2013){Rusli}, {Thomas}, {Saglia}, {Fabricius},
  {Erwin}, {Bender}, {Nowak}, {Lee}, {Riffeser}, \& {Sharp}}]{Rusli2013}
{Rusli}, S.~P., {Thomas}, J., {Saglia}, R.~P., {et~al.} 2013, \aj, 146, 45

\bibitem[{{Rybicki}(1987)}]{Rybicki1987}
{Rybicki}, G.~B. 1987, in IAU Symposium, Vol. 127, Structure and Dynamics of
  Elliptical Galaxies, ed. P.~T. {de Zeeuw}, 397

\bibitem[{{Saglia} {et~al.}(2016){Saglia}, {Opitsch}, {Erwin}, {Thomas},
  {Beifiori}, {Fabricius}, {Mazzalay}, {Nowak}, {Rusli}, \&
  {Bender}}]{Saglia2016}
{Saglia}, R.~P., {Opitsch}, M., {Erwin}, P., {et~al.} 2016, \apj, 818, 47

\bibitem[{{Salpeter}(1955)}]{Salpeter1955}
{Salpeter}, E.~E. 1955, \apj, 121, 161

\bibitem[{{S{\'a}nchez} {et~al.}(2012){S{\'a}nchez}, {Kennicutt}, {Gil de Paz},
  {van de Ven}, {V{\'{\i}}lchez}, {Wisotzki}, {Walcher}, {Mast}, {Aguerri},
  {Albiol-P{\'e}rez}, {Alonso-Herrero}, {Alves}, {Bakos}, {Bart{\'a}kov{\'a}},
  {Bland-Hawthorn}, {Boselli}, {Bomans}, {Castillo-Morales}, {Cortijo-Ferrero},
  {de Lorenzo-C{\'a}ceres}, {Del Olmo}, {Dettmar}, {D{\'{\i}}az}, {Ellis},
  {Falc{\'o}n-Barroso}, {Flores}, {Gallazzi}, {Garc{\'{\i}}a-Lorenzo},
  {Gonz{\'a}lez Delgado}, {Gruel}, {Haines}, {Hao}, {Husemann},
  {Igl{\'e}sias-P{\'a}ramo}, {Jahnke}, {Johnson}, {Jungwiert}, {Kalinova},
  {Kehrig}, {Kupko}, {L{\'o}pez-S{\'a}nchez}, {Lyubenova}, {Marino},
  {M{\'a}rmol-Queralt{\'o}}, {M{\'a}rquez}, {Masegosa}, {Meidt},
  {Mendez-Abreu}, {Monreal-Ibero}, {Montijo}, {Mour{\~a}o}, {Palacios-Navarro},
  {Papaderos}, {Pasquali}, {Peletier}, {P{\'e}rez}, {P{\'e}rez}, {Quirrenbach},
  {Rela{\~n}o}, {Rosales-Ortega}, {Roth}, {Ruiz-Lara},
  {S{\'a}nchez-Bl{\'a}zquez}, {Sengupta}, {Singh}, {Stanishev}, {Trager},
  {Vazdekis}, {Viironen}, {Wild}, {Zibetti}, \& {Ziegler}}]{Sanchez2012}
{S{\'a}nchez}, S.~F., {Kennicutt}, R.~C., {Gil de Paz}, A., {et~al.} 2012,
  \aap, 538, A8

\bibitem[{{S{\'a}nchez-Bl{\'a}zquez} {et~al.}(2006){S{\'a}nchez-Bl{\'a}zquez},
  {Peletier}, {Jim{\'e}nez-Vicente}, {Cardiel}, {Cenarro},
  {Falc{\'o}n-Barroso}, {Gorgas}, {Selam}, \&
  {Vazdekis}}]{Sanchez-Blazquez2006}
{S{\'a}nchez-Bl{\'a}zquez}, P., {Peletier}, R.~F., {Jim{\'e}nez-Vicente}, J.,
  {et~al.} 2006, \mnras, 371, 703

\bibitem[{{Sarzi} {et~al.}(2001){Sarzi}, {Rix}, {Shields}, {Rudnick}, {Ho},
  {McIntosh}, {Filippenko}, \& {Sargent}}]{Sarzi2001}
{Sarzi}, M., {Rix}, H.-W., {Shields}, J.~C., {et~al.} 2001, \apj, 550, 65

\bibitem[{{Schulze} \& {Gebhardt}(2011)}]{Schulze2011a}
{Schulze}, A. \& {Gebhardt}, K. 2011, \apj, 729, 21

\bibitem[{{Schwarzschild}(1979)}]{Schwarzschild1979}
{Schwarzschild}, M. 1979, \apj, 232, 236

\bibitem[{{Scott} {et~al.}(2013){Scott}, {Cappellari}, {Davies}, {Kleijn},
  {Bois}, {Alatalo}, {Blitz}, {Bournaud}, {Bureau}, {Crocker}, {Davis}, {de
  Zeeuw}, {Duc}, {Emsellem}, {Khochfar}, {Krajnovi{\'c}}, {Kuntschner},
  {McDermid}, {Morganti}, {Naab}, {Oosterloo}, {Sarzi}, {Serra}, {Weijmans}, \&
  {Young}}]{Scott2013}
{Scott}, N., {Cappellari}, M., {Davies}, R.~L., {et~al.} 2013, \mnras, 432,
  1894

\bibitem[{{Seth} {et~al.}(2014){Seth}, {van den Bosch}, {Mieske}, {Baumgardt},
  {Brok}, {Strader}, {Neumayer}, {Chilingarian}, {Hilker}, {McDermid},
  {Spitler}, {Brodie}, {Frank}, \& {Walsh}}]{Seth2014}
{Seth}, A.~C., {van den Bosch}, R., {Mieske}, S., {et~al.} 2014, \nat, 513, 398

\bibitem[{{Shetty} \& {Cappellari}(2015)}]{Shetty2015}
{Shetty}, S. \& {Cappellari}, M. 2015, \mnras, 454, 1332

\bibitem[{{Skrutskie} {et~al.}(2006){Skrutskie}, {Cutri}, {Stiening},
  {Weinberg}, {Schneider}, {Carpenter}, {Beichman}, {Capps}, {Chester},
  {Elias}, {Huchra}, {Liebert}, {Lonsdale}, {Monet}, {Price}, {Seitzer},
  {Jarrett}, {Kirkpatrick}, {Gizis}, {Howard}, {Evans}, {Fowler}, {Fullmer},
  {Hurt}, {Light}, {Kopan}, {Marsh}, {McCallon}, {Tam}, {Van Dyk}, \&
  {Wheelock}}]{Skrutskie2006}
{Skrutskie}, M.~F., {Cutri}, R.~M., {Stiening}, R., {et~al.} 2006, \aj, 131,
  1163

\bibitem[{{Thomas} {et~al.}(2016){Thomas}, {Ma}, {McConnell}, {Greene},
  {Blakeslee}, \& {Janish}}]{Thomas2016}
{Thomas}, J., {Ma}, C.-P., {McConnell}, N.~J., {et~al.} 2016, \nat, 532, 340

\bibitem[{{Tremaine} {et~al.}(2002){Tremaine}, {Gebhardt}, {Bender}, {Bower},
  {Dressler}, {Faber}, {Filippenko}, {Green}, {Grillmair}, {Ho}, {Kormendy},
  {Lauer}, {Magorrian}, {Pinkney}, \& {Richstone}}]{Tremaine2002}
{Tremaine}, S., {Gebhardt}, K., {Bender}, R., {et~al.} 2002, \apj, 574, 740

\bibitem[{{Valdes} {et~al.}(2004){Valdes}, {Gupta}, {Rose}, {Singh}, \&
  {Bell}}]{Valdes2004}
{Valdes}, F., {Gupta}, R., {Rose}, J.~A., {Singh}, H.~P., \& {Bell}, D.~J.
  2004, \apjs, 152, 251

\bibitem[{{Vallejo} {et~al.}(2002){Vallejo}, {Braine}, \&
  {Baudry}}]{Vallejo2002}
{Vallejo}, O., {Braine}, J., \& {Baudry}, A. 2002, \aap, 387, 429

\bibitem[{{Vallejo} {et~al.}(2003){Vallejo}, {Braine}, \&
  {Baudry}}]{Vallejo2003}
{Vallejo}, O., {Braine}, J., \& {Baudry}, A. 2003, \apss, 284, 715

\bibitem[{{Valluri} {et~al.}(2005){Valluri}, {Ferrarese}, {Merritt}, \&
  {Joseph}}]{Valluri2005}
{Valluri}, M., {Ferrarese}, L., {Merritt}, D., \& {Joseph}, C.~L. 2005, \apj,
  628, 137

\bibitem[{{van den Bosch}(2016)}]{VandenBosch2016}
{van den Bosch}, R. 2016, ArXiv e-prints

\bibitem[{{van den Bosch} {et~al.}(2012){van den Bosch}, {Gebhardt},
  {G{\"u}ltekin}, {van de Ven}, {van der Wel}, \& {Walsh}}]{vandenBosch2012}
{van den Bosch}, R.~C.~E., {Gebhardt}, K., {G{\"u}ltekin}, K., {et~al.} 2012,
  \nat, 491, 729

\bibitem[{{van der Marel} {et~al.}(1998){van der Marel}, {Cretton}, {de Zeeuw},
  \& {Rix}}]{vanderMarel1998}
{van der Marel}, R.~P., {Cretton}, N., {de Zeeuw}, P.~T., \& {Rix}, H.-W. 1998,
  \apj, 493, 613

\bibitem[{{van der Marel} \& {Franx}(1993)}]{vanderMarel1993}
{van der Marel}, R.~P. \& {Franx}, M. 1993, \apj, 407, 525

\bibitem[{{Vazdekis} {et~al.}(2010){Vazdekis}, {S{\'a}nchez-Bl{\'a}zquez},
  {Falc{\'o}n-Barroso}, {Cenarro}, {Beasley}, {Cardiel}, {Gorgas}, \&
  {Peletier}}]{Vazdekis2010}
{Vazdekis}, A., {S{\'a}nchez-Bl{\'a}zquez}, P., {Falc{\'o}n-Barroso}, J.,
  {et~al.} 2010, \mnras, 404, 1639

\bibitem[{{Verdoes Kleijn} {et~al.}(2002){Verdoes Kleijn}, {van der Marel}, {de
  Zeeuw}, {Noel-Storr}, \& {Baum}}]{VerdoesKleijn2002}
{Verdoes Kleijn}, G.~A., {van der Marel}, R.~P., {de Zeeuw}, P.~T.,
  {Noel-Storr}, J., \& {Baum}, S.~A. 2002, \aj, 124, 2524

\bibitem[{{Vittorini} {et~al.}(2005){Vittorini}, {Shankar}, \&
  {Cavaliere}}]{Vittorini2005}
{Vittorini}, V., {Shankar}, F., \& {Cavaliere}, A. 2005, \mnras, 363, 1376

\bibitem[{{Walsh} {et~al.}(2015){Walsh}, {van den Bosch}, {Gebhardt},
  {Yildirim}, {G{\"u}ltekin}, {Husemann}, \& {Richstone}}]{Walsh2015}
{Walsh}, J.~L., {van den Bosch}, R.~C.~E., {Gebhardt}, K., {et~al.} 2015, \apj,
  808, 183

\bibitem[{{Walsh} {et~al.}(2016){Walsh}, {van den Bosch}, {Gebhardt},
  {Y{\i}ld{\i}r{\i}m}, {Richstone}, {G{\"u}ltekin}, \& {Husemann}}]{Walsh2016}
{Walsh}, J.~L., {van den Bosch}, R.~C.~E., {Gebhardt}, K., {et~al.} 2016, \apj,
  817, 2

\bibitem[{{Winge} {et~al.}(2009){Winge}, {Riffel}, \&
  {Storchi-Bergmann}}]{Winge2009}
{Winge}, C., {Riffel}, R.~A., \& {Storchi-Bergmann}, T. 2009, \apjs, 185, 186

\bibitem[{{Wong} {et~al.}(2004){Wong}, {Blitz}, \& {Bosma}}]{Wong2004}
{Wong}, T., {Blitz}, L., \& {Bosma}, A. 2004, \apj, 605, 183

\end{thebibliography}
\end{document}